\documentclass[conference]{IEEEtran}
\IEEEoverridecommandlockouts
\usepackage{cite}
\usepackage{amsmath,amssymb,amsfonts}
\usepackage{algorithmic}
\usepackage[ruled,vlined]{algorithm2e}
\usepackage{subcaption}
\usepackage[hidelinks]{hyperref}
\usepackage{float}
\usepackage{booktabs}
\usepackage{graphicx}
\usepackage{textcomp}
\usepackage{placeins}
\usepackage{makecell}
\usepackage{xcolor}
\usepackage{tabularx}
\usepackage{longtable}
\def\BibTeX{{\rm B\kern-.05em{\sc i\kern-.025em b}\kern-.08em
    T\kern-.1667em\lower.7ex\hbox{E}\kern-.125emX}}
\usepackage{array}
\newcolumntype{P}[1]{>{\raggedright\arraybackslash}p{#1}}  

\begin{document}

\title{BitFlipScope: Scalable Fault Localization and Recovery for Bit-Flip Corruptions in LLMs \\}

\author{\IEEEauthorblockN{1\textsuperscript{st} Muhammad Zeeshan Karamat}
\IEEEauthorblockA{\textit{Bradly Dept. of ECE} \\
\textit{Virginia Tech}\\
Blacksburg, Virginia, USA \\
mzeeshan@vt.edu}
\and
\IEEEauthorblockN{2\textsuperscript{nd} Sadman Saif}
\IEEEauthorblockA{\textit{Bradly Dept. of ECE} \\
\textit{Virginia Tech}\\
Blacksburg, Virginia, USA \\
sadmansaif@vt.edu}
\and
\IEEEauthorblockN{3\textsuperscript{rd} Christiana Chamon}
\IEEEauthorblockA{\textit{Bradly Dept. of ECE} \\
\textit{Virginia Tech}\\
Blacksburg, Virginia, USA \\
ccgarcia@vt.edu}
}

% \author{
% \IEEEauthorblockN{Anonymous Author(s)}
% \IEEEauthorblockA{
% Anonymous Institution\\
% City, Country\\
% email@anonymous.com
% }
% }

% \author{
% \IEEEauthorblockA{\vspace{1em}}
% \IEEEauthorblockN{Anonymous Author(s)}
% \IEEEauthorblockA{\vspace{1em}}
% }

\newcommand{\newedits}[1]{ \color{black} #1  \color{black}}
\maketitle

\begin{abstract}
Large Language Models (LLMs) deployed in practical and safety-critical settings are increasingly susceptible to bit-flip faults caused by hardware degradation, cosmic radiation, or deliberate fault-injection attacks such as Rowhammer. These faults silently corrupt internal parameters and can lead to unpredictable or dangerous model behavior. \textbf{Localizing these corruptions is essential:} without identifying the affected region, it is impossible to diagnose the source of degradation, apply targeted corrective measures, or restore model functionality without resorting to costly fine-tuning or full retraining. This work introduces \textbf{BitFlipScope}, a scalable, software-based framework for identifying fault-affected regions within transformer architectures under two deployment scenarios. When a clean reference model is available, BitFlipScope performs differential analysis of outputs, hidden states, and internal activations for detecting anomalous behavior indicative of corruption to pinpoint or localize faults. When no reference model exists, it uses residual-path perturbation and loss-sensitivity profiling to infer the fault-impacted region directly from the corrupted model. In both settings, the framework not only enables effective fault diagnosis but also supports lightweight performance recovery without fine-tuning, offering a practical path to restoring corrupted models. Together, these capabilities make BitFlipScope an important step toward trustworthy, fault-resilient LLM deployment in hardware-prone and adversarial environments.

\end{abstract}


\section{Introduction}
\label{sec:Intro}
\begin{figure*}[t]
    \centering
    \includegraphics[width=0.95\textwidth]{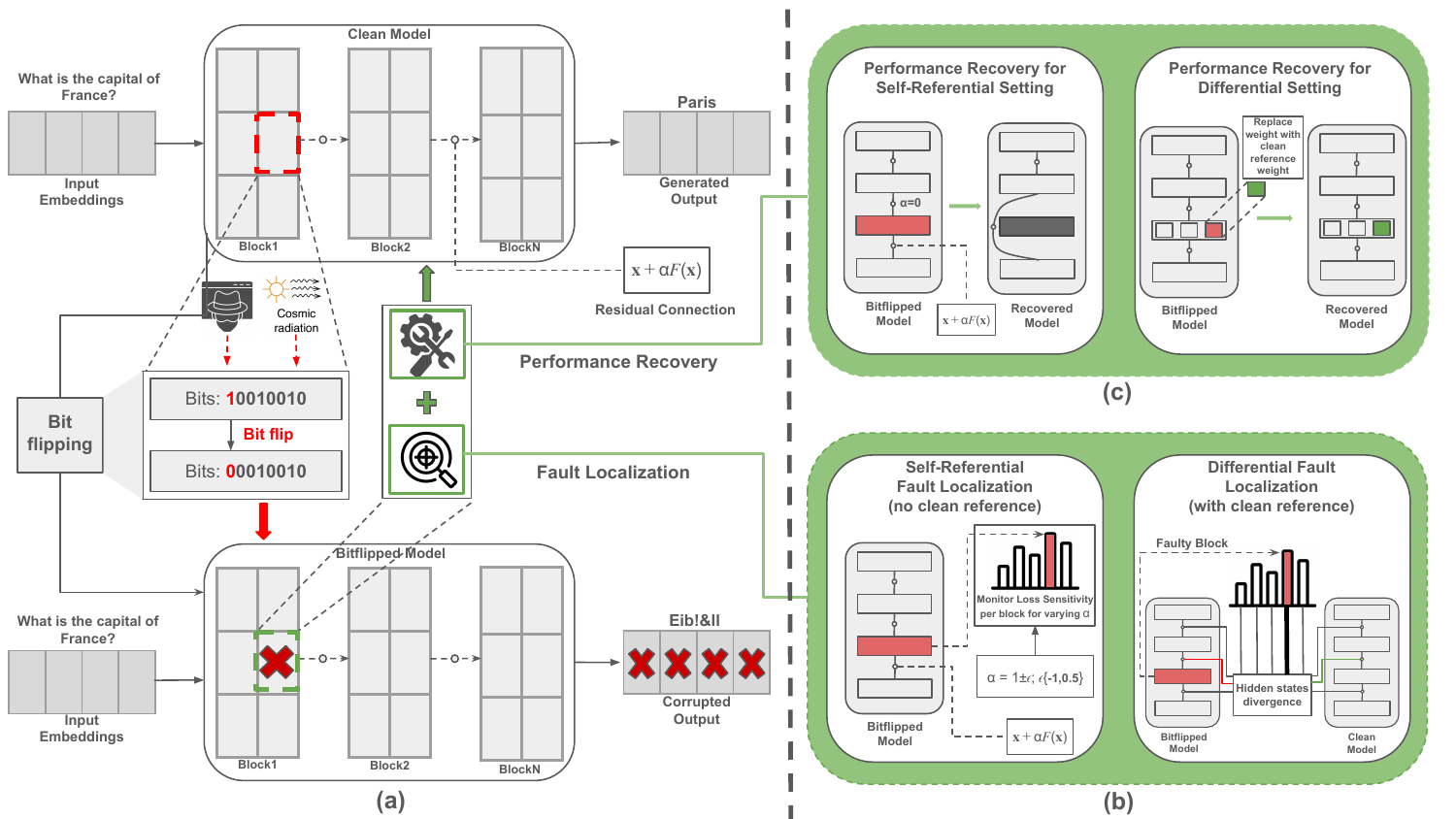}
\caption{
Overview of the \textbf{BitFlipScope} framework for detecting and mitigating bit-flip faults in LLMs. 
\textbf{(a)} A single bit-flip arising from hardware faults or attack corrupts a transformer block and degrades the model’s output. 
\textbf{(b)} Fault localization is performed using two approaches: \emph{self-referential} analysis (left), which identifies abnormal loss sensitivity under residual scaling, and \emph{differential} analysis (right), which detects hidden-state divergence using a clean reference model. 
\textbf{(c)} Once the faulty block is identified, lightweight recovery mechanisms reduce or correct its influence, enabling performance restoration without fine-tuning.
}

    \label{fig:overview}
\end{figure*}
% The impeccable usability and convenience of Large Language Models (LLMs) have facilitated their widespread adoption in various fields, including education, finance, and medicine. Working with billions of parameters, LLM offers natural language processing, targeted output generation, pattern recognition, and desired information extraction \cite{Raza_Jahangir_Riaz_Saeed_Sattar_2025}. Given our reliance on LLM outputs across various applications, ensuring the accuracy of these outputs is paramount. Erroneous output from the LLMs in critical systems could lead to severe or catastrophic consequences. Misinformation, hallucination, and trustworthiness are common security concerns and vulnerabilities of LLMs. Gradient-based bit-flip attacks, adversarial attacks, prompt injection, hardware-based attacks, etc., can trigger substantial performance degradation in LLMs' output \cite{SPCLLMs}.

The impeccable usability and convenience of Large Language Models (LLMs) have facilitated their widespread adoption in various fields, including education, finance, and medicine \cite{liang2025widespread, xu2024large, chen2024a, zhou2023survey}. Working with billions of parameters, LLM offers natural language processing, targeted output generation, pattern recognition, and desired information extraction \cite{zhao2023survey}. Given our reliance on LLM outputs across various applications, ensuring the accuracy of these outputs is paramount. Erroneous output from the LLMs in critical systems could lead to severe or catastrophic consequences. Misinformation, hallucination, and trustworthiness are common security concerns and vulnerabilities of LLMs \cite{bender2021stochastic, bommasani2021opportunities}. Gradient-based bit-flip attacks, adversarial attacks, prompt injection, hardware-based attacks, etc., can trigger substantial performance degradation in LLMs' output \cite{SPCLLMs, BitflipDNN, GenBFA}.

Among these attack strategies against LLMs, the bit-flip attack is the most subtle yet damaging, as this strategy targets a small number of the most impactful weights and modifies a single bit for each of those selected weights. This silent corruption of weights causes an unprecedented drop in accuracy for the model's output. This focused manipulation of 10-128 weights among billions of weights can cause a loss of accuracy of 73\% on the GPT-2 32-bit model\cite{LLWRA}. LLMs are vulnerable to this type of attack through hardware induction as well. The MobileNet-2 model's performance can be compromised to 75\% accuracy loss through deterministic bitflips of only 2 weights, leveraging the Rowhammer vulnerability \cite{Deephammer}.

To initiate countermeasures for reinstating the model's performance, debugging or fault recovery, it is essential to pinpoint exactly which weights have been manipulated by the attack. Existing attack or defense methods lack fine-grained localization as they either quantify global performance degradation or rely on exhaustive parameter comparisons, both of which are infeasible for large-scale models. Given that modern LLMs contain billions of parameters distributed across hundreds of layers, scalable fault localization is an unsolved challenge. In this work, we investigate: how can we design an ideal fault localization tool for LLMs that has \newedits{(1) granularity, enabling block‑level localization in all deployment settings and, where a clean reference is available, refinement down to the layer, weight, and bit level}; (2) efficiency, minimizing computational overhead compared to brute-force methods; and (3) scalability, operating feasibly on billion-parameter models without retraining or intrusive architectural modification?

\newedits{Precise and efficient identification of manipulated bits requires tailored strategies, which differ depending on the availability of a clean reference model.} With access to a clean model and an attacked model for differential settings, a comparative parametric analysis can be performed. \newedits{This allows for the detection of the layer, block, weights, and bits affected by fault-induced corruption.} In most real-world deployments, the clean reference model is not accessible, which complicates pinpointing the source of faults.\newedits{Therefore, it is essential to develop a self-referential method that directly localizes block-level bit-flip faults from a fault-injected model without requiring baseline model comparisons.}

In this paper, we present \textbf{BitFlipScope}, a unified, software-based framework for efficiently localizing and mitigating bit-flip faults in large language models across both differential and self-referential deployment settings. The main contributions of this work are:

\begin{itemize}
    \item \textbf{Dual-setting fault localization:} \newedits{We propose a unified framework that localizes bit‑flip faults at the block level in both self‑referential and differential settings, and further refines faults down to the layer, weight, and bit levels in the differential setting, where a clean reference model is available.}

    \item \textbf{Lightweight performance recovery:} We demonstrate practical, fine-tuning-free mitigation strategies, including targeted parameter restoration in the differential setting and scaling-based attenuation in the self-referential setting.

    \item \textbf{Scalable evaluation:} We validate BitFlipScope across multiple model sizes and corruption sites, showing consistent localization accuracy and substantial performance recovery.
\end{itemize}

\noindent Together, these contributions advance the reliability and security of LLM deployments by enabling practical fault localization and recovery under both constrained and fully informed operating environments.

\section{Background}
\label{sec:background}
This section provides an overview of bit flip attacks on neural networks and LLMs, along with related detection, localization, and recovery approaches.

\subsection{Bit-flip Attacks}

\newedits{Before presenting localization and recovery mechanisms, we briefly review representative bit flip attack strategies.} Most attacks target vulnerable regions of a model using search strategies that identify parameters whose corruption produces maximal impact.

For hardware based attacks on deep neural networks, Progressive Bit Search (PBS) ranks vulnerable bits within and across layers using gradient information, and corrupts them through the Row Hammer vulnerability. Using this approach, \cite{BitflipDNN} reduced the accuracy of a ResNet 18 model from 69.8\% to 0.1\% by flipping only 13 bits. Similarly, \cite{Deephammer} exploits DRAM Row Hammer to corrupt quantized DNN weights using a flip aware search that ranks vulnerable bits and ensures precise DRAM bit flips. Experiments across 11 architectures show that 2 to 24 bit flips can compromise a model, with ResNet requiring over 20 flips while MobileNet V2 suffers more than 75\% accuracy loss with only 2 flips.

Another physical fault injection method, LLWRA \cite{LLWRA}, targets LLMs through Row Hammer attacks on Page Frame Numbers. The ReBlock algorithm identifies the most influential block of 128 weights using gradient analysis and replaces it with a block requiring minimal bit flips. This iterative block replacement reduces GPT 2 accuracy on WikiText from 78\% to below 5\% within 9 iterations.

For software based attacks, GenBFA \cite{GenBFA} degrades performance of the LLaMA3 8B Instruct 8 bit quantized model from 67.3\% to 0\% using only 3 bit flips. The attack first identifies a critical layer through layer wise sensitivity analysis, then perturbs subsets of weights within that layer to locate the most vulnerable parameters and minimize the number of required bit flips.

Both hardware and software based attacks demonstrate the severe vulnerability of neural networks and LLMs to small numbers of bit flips.

\subsection{Localization of the Faults}
In all of the attack methods, the most impactful layers, blocks, weights, or bits selection is evident to make the attack efficient and subtle. Localization of the most vulnerable part of the LLM to attack is well-explored which is from an attacker's perspective. In this paper, we examine post-attack fault localization in a deployed LLM model from the perspective of the entity responsible for recovery. In \cite{BFADetection}, a bit-flip attack detection scheme is demonstrated. In this article, a bit-flip encoding-based detection method is introduced. First, a sensitivity analysis of the quantized neural network's weights determines the most critical ones. These selected weights are encoded with a secret key into a binary detection code. These sensitive weights are monitored by continuously computing the Hamming distance between the current detection code and the securely stored code. Large Hamming distance dictates that there has been an attack.

But this scheme does not pinpoint exact bits or weights that have been manipulated. Whether a model is attacked or not can be detected by analyzing the output of that model. To the best of our knowledge, no prior work provides a practical, scalable tool to localize bit-flip faults in LLMs at block/layer/weight/bit granularity.\newedits{This paper presents a scalable framework that enables block-level fault localization in both differential and self-referential settings. While fine-grained localization (to layer and bit-level) is achievable in the differential setting, the self-referential case supports only block-level identification, potentially enabling the development of new recovery strategies for compromised models.}

\subsection{Performance Recovery Techniques After Bit-flip Attack}
\label{subsec:Back-recovery}
With the goal of introducing a resilient transformer-based model, in \cite{FAR}, a unique architecture is demonstrated called Forget and Rewire. This approach finds out ineffective weight parameters using sensitivity analysis based on the gradients of the weights. After identifying those dead weights, this algorithm forgets them and rewrites those configurations of inputs with a division factor that dilutes the gradient of a significant weight by half. From the attackers' perspective, the weight rankings change, and the attackers need to flip more bits than before to do the same performance degradation. With this scheme implemented, the model's performance drop no more than 2\%. Another approach called NeuroPots \cite{NeuroPots} introduces a proactive defense mechanism for neural networks. This mechanism adds honey neurons, which are enhanced activation values with an expanding coefficient that puts these neurons up in the weight ranking to attack. These particular neurons will be monitored, and if the attack is detected, then securely stored clean honey weights will be replaced by clean ones. The model's accuracy can be restored to 90\% after mitigation. \newedits{Another method explored in the deep neural network domain \cite{NAPER} detects faults by verifying a layer-wise relation between base and redundant model weights, using quick sum checks followed by checksums to isolate corruption. Recovery inverts this relation to reconstruct faulty parameters, with a scheduler managing the process to maintain inference. While effective for compact models, its reliance on per-layer comparisons, redundant models, and layer slack makes it unsuitable for LLMs due to their scale, memory demands, and tightly coupled transformer layers.} In \cite{ZeroingMethod}, a fault tolerance technique is showcased for Vision Transformers where the Least Significant Bit of all parameters is replaced with a parity bit and forced to have an even parity. This works as an error detection system when a bit is flipped, then in the parity check, a flag is raised as an attack. The affected parameter is zeroed out. This mechanism can keep ViT's accuracy above 85\% with a threshold Bit Error Rate of 2e-5.\newedits{In the setting of language models, LM-Fix \cite{lmfix} introduces a bit flip detection scheme that feeds a fixed test vector through the model and compares the final layer tensor with a stored reference to detect corruption. For recovery, it flushes caches, identifies the faulty layer by comparing intermediate activations, locates corrupted rows and columns through a rotation-based search, and restores weights by solving a linear system using redundancy buffers, avoiding a full model reload. However, it fails to detect low-impact Silent Safe Bit Flips and relies on a single fixed test vector that can be exploited by adaptive attacks. The fixed vector may not activate certain neurons or weight paths. Moreover, the method assumes linearity even in nonlinear layers and introduces memory overhead of up to 5\%.}\\
Aside from LM-Fix, most fault detection and recovery techniques are designed for neural networks or small-scale transformers. Large language models contain billions of parameters, and bit flip attacks on them produce subtle but progressively degrading effects. Consequently, a gap remains in localizing bit flip attacks and recovering using the attacked LLM itself. To address this gap, BitFlipScope introduces recovery mechanisms in both differential and self-referential settings, enabling post attack localization and recovery of bit flips without retraining, auxiliary neurons, or pre-embedded fault tolerant structures.

\section{Methodology}
\label{sec:methodology}

The overall BitFlipScope workflow is illustrated in Fig.~\ref{fig:overview}. In practice, bit-flip faults may arise under two distinct deployment scenarios: (i) a \emph{differential} setting, where a clean reference model is available, and (ii) a \emph{self-referential} setting, where only the corrupted model can be accessed. These scenarios differ fundamentally in the information available for diagnosis and motivate the two complementary localization paths in our framework.

Accordingly, this section is organized into three parts. We first formalize the two fault-localization settings. We then describe the localization methodology for each case. Finally, we present the lightweight performance recovery mechanisms that use the localized fault information to restore model behavior without fine-tuning.

\subsection{Fault Localization Settings}

\paragraph{Differential Setting}
In some deployments, a clean reference model may be preserved in secure storage, mirrored across nodes, or retrievable from a verified checkpoint. This setting allows behavioral comparison between the corrupted and clean models. \newedits{The primary challenge involves achieving efficient comparisons, avoiding computationally impractical parameter-level assessments across an extensive weight population.}

\newedits{Naively, one might assume that if a clean clone of the model exists, the simplest solution is a full parameter comparison followed by direct replacement of mismatched weights. In practice, this is often infeasible for LLMs: moving or restoring large model states is frequently limited by storage and network I/O rather than arithmetic throughput. In addition, practical checkpoints may be stored in incremental or chunked formats rather than as a single flat parameter snapshot, and checkpoint artifacts can themselves be corrupted, making naive full weight replacement unreliable \cite{IncrCP, ROJAS2024104879, rajbhandari2020zero, narayanan2021efficient}. BitFlipScope sidesteps these constraints by first localizing faults at the block and layer level using activation divergence, and only then applying hash‑guided comparison to a small subset of tensors. This staged design reduces the number of bytes that must be read and compared by orders of magnitude while tolerating partially degraded or compressed references.}

\paragraph{Self-Referential Setting}
In many realistic scenarios edge deployments, memory-constrained systems, or models updated in-place—only the corrupted model is available. No checkpoints or clean replicas can be accessed, and the system must diagnose faults from the corrupted model alone. This scenario is significantly more challenging, because no ground-truth behavior is available for comparison. Our method addresses this by exploiting structural properties of transformer architectures to infer abnormal computation patterns without any external reference.

\subsection{Self-Referential Fault Localization}
\label{subsec:self-ref_FL}
Transformer blocks contribute to the forward computation through residual pathways. Assuming the input to the $i$-th transformer block is $\mathbf{h}_i$, the block output can be expressed as a composition of its attention and feed-forward components, each equipped with a residual pathway. Following the standard transformer formulation~\cite{llama3meta2024}, the block computes:
\[
\mathbf{h}_{i+1} 
= \mathbf{h}_i 
+ \underbrace{\text{ATTN}_i(\mathbf{h}_i)}_{\text{self-attention}} 
+ \underbrace{\text{MLP}_i\!\left(\mathbf{h}_i + \text{ATTN}_i(\mathbf{h}_i)\right)}_{\text{feed-forward}},
\]
where $\text{ATTN}_i(\cdot)$ and $\text{MLP}_i(\cdot)$ denote the attention and feed-forward transformations, respectively. Both submodules contribute additively through residual connections, forming the dominant pathway through which information propagates across the depth of the model.

Because residual connections directly control the influence of each block on subsequent computations, scaling the residual contribution of block $i$ by a factor $\alpha_i$ effectively modulates the distribution of its output:
\[
\mathbf{h}_{i+1} = \mathbf{h}_i + \alpha_i F_i(\mathbf{h}_i).
\]

where $F_i(\cdot)$ is the block’s transformation (self-attention and MLP components). When $\alpha_i > 1$, the block’s contribution is amplified, increasing the magnitude of its output and correspondingly affecting the input to block $i{+}1$ during inference. This amplification changes the distribution of the hidden-state vector and effectively strengthens the block’s role in the autoregressive token-generation process. Conversely, when $\alpha_i < 1$, the block’s influence is diminished, reducing its contribution to the model’s forward computation. Applying such scaling across blocks similarly increases or decreases their collective effect on the resulting output distribution. This controllable adjustment of the block's output distribution forms the foundation of our self-referential localization approach.

% Our approach, summarized in Algorithm~\ref{alg:selfref} leverages this structure to isolate corrupted regions by perturbing the residual scaling and examining the induced change in model loss. 

% For block $i$, the residual update is given by:
% \[
% \mathbf{h}_{i+1} = \mathbf{h}_{i} + \alpha_i \cdot F_i(\mathbf{h}_{i}),
% \]
% where $\mathbf{h}_i$ is the block input, $F_i(\cdot)$ is the block’s transformation (self-attention and MLP components), and $\alpha_i$ is a tunable scaling parameter. In the unmodified model, $\alpha_i = 1$ for all blocks.

\paragraph{Residual Perturbation}
To identify corrupted blocks, we probe the model’s behavior under controlled deviations of the scaling parameter $\alpha_i$ from its nominal value of $1$, following the procedure summarized in Algorithm~\ref{alg:selfref}. For each block, we evaluate the model after slightly increasing its residual contribution (\textit{scale-up}: $\alpha_i = 1 + \epsilon$) and slightly decreasing it (\textit{scale-down}: $\alpha_i = 1 - \epsilon$). In a healthy block, such small adjustments produce only minor and predictable changes in the model’s output distribution. However, when a block’s internal parameters have been corrupted by a bit-flip, these adjustments expose its abnormal influence: increasing $\alpha_i$ strengthens the effect of the corrupted transformation, while decreasing $\alpha_i$ weakens it. This asymmetric behavioral response provides the foundation for detecting the presence of a faulty block.

\paragraph{Loss Change Metric}
To quantify how each block responds to residual scaling, we measure the change in model loss induced by different values of $\alpha_i$. For block $\ell$, we define the loss change as:
\[
\Delta \mathrm{Loss}(\ell, \alpha)
    = \mathrm{Loss}(\ell, \alpha) - \mathrm{Loss}_{\mathrm{base}},
\]
where $\mathrm{Loss}_{\mathrm{base}}$ denotes the loss under the unscaled model ($\alpha = 1$). This quantity captures how strongly the model’s output distribution shifts when the influence of block $\ell$ is amplified or attenuated. Blocks whose computations are intact yield small and consistent $\Delta \mathrm{Loss}$ values, whereas corrupted blocks exhibit disproportionately large deviations. This metric provides the quantitative foundation for identifying abnormal block behavior.

\paragraph{Scaling Values}
Residual scaling factors are selected based on an empirical $\alpha$-sweep that measures the sensitivity of the loss to residual modulation. In practice, we use the discrete set
\[
\alpha \in \{0.6,\,0.7,\,0.8,\,0.9,\,1.1,\,1.2,\,1.3,\,1.4\},
\]
which probes both attenuation and amplification around the nominal value $\alpha=1$.\newedits{The empirical procedure used to determine this range is described in Appendix~\ref{appendix:alpha_sweep}}.

\paragraph{Sensitivity Metric}
After computing the loss change for each block across the selected scaling values, we aggregate these responses into a single score that reflects how strongly each block reacts to residual modulation. Let $\mathcal{A}$ denote the set of scaling factors chosen from the $\alpha$-sweep. For block $\ell$, we define the Block Sensitivity Score (BSS) as:
\[
\mathrm{BSS}(\ell)
= 
\sum_{\alpha \in \mathcal{A}}
\left|\Delta \mathrm{Loss}(\ell, \alpha)\right|,
\]
which can be expanded using the loss-change definition as:
\[
\mathrm{BSS}(\ell)
=
\sum_{\alpha \in \mathcal{A}}
\left|\,\mathrm{Loss}(\ell, \alpha)
- \mathrm{Loss}_{\mathrm{base}}\,\right|
\]

Blocks whose computations are intact exhibit small and consistent sensitivity scores because their loss remains largely stable across the selected $\alpha$ values. In contrast, a corrupted block yields a distinctly higher score, reflecting the amplified effect of scaling on its abnormal output distribution. We therefore identify the block with the highest BSS as the anomalous and thus likely faulty block. This metric is lightweight, fully gradient-free, and exploits predictable error propagation patterns induced by bit-level corruption.

\newedits{
\paragraph{Robust Sensitivity-Based Detection}
\label{para:robust_detection}

Given the Block Sensitivity Scores (BSS) $\{\mathrm{BSS}(\ell)\}$ computed across all layers, we detect anomalous blocks using a robust normalization scheme. We estimate the median and Median Absolute Deviation (MAD) of the BSS distribution~\cite{huber1981robust} and compute a robust z-score for each layer. This normalization evaluates sensitivity relative to typical layer behavior, mitigating the influence of extreme outliers and ensuring scale invariance across models and datasets.

A layer is declared anomalous if its robust z-score exceeds a threshold $\tau$. The threshold is derived from extreme-value considerations and scales logarithmically with the number of layers, providing principled control of false positives. The statistical justification for the robust normalization and threshold selection is provided in Appendix~\ref{appendix:robust_detection}.

To enable multi-block detection, we adopt an iterative procedure. At each iteration, the layer with the largest robust z-score is identified. If its score exceeds $\tau$, the layer is marked as faulty and its residual contribution is neutralized. Sensitivity scores are then recomputed on the modified model. The procedure terminates when no layer exceeds the detection threshold.

This iterative neutralization prevents dominant faults from masking weaker ones and enables reliable localization of multiple corrupted blocks within the network.

\begin{algorithm}[h]
\newedits{
\caption{Self-Referential Fault Localization via Iterative Residual Neutralization}
\label{alg:selfref}
\KwIn{Faulty model $M$, Input set $\mathcal{X}$, Scaling set $\mathcal{A}$, Threshold $\tau$}
\KwOut{Detected faulty block set $\mathcal{F}$}

\BlankLine
Initialize $\mathcal{F} \leftarrow \emptyset$\;

\While{true}{

    \ForEach{block $\ell$ in the transformer}{
        Initialize $\mathrm{BSS}(\ell) \leftarrow 0$\;

        \ForEach{$\alpha \in \mathcal{A}$}{
            Temporarily scale residual of block $\ell$ by $\alpha$\;
            Compute loss $\mathcal{L}(\ell,\alpha)$ on $\mathcal{X}$\;
            $\mathrm{BSS}(\ell) \leftarrow \mathrm{BSS}(\ell) + |\mathcal{L}(\ell,\alpha) - \mathcal{L}_{\text{base}}|$\;
        }

        Reset residual scaling to $\alpha = 1$\;
    }

    Compute median and MAD of $\{\mathrm{BSS}(\ell)\}$\;
    Compute robust z-scores $z(\ell)$ for all blocks\;

    $\ell^* \leftarrow \arg\max_\ell z(\ell)$\;

    \If{$z(\ell^*) < \tau$}{
        break\;
    }

    $\mathcal{F} \leftarrow \mathcal{F} \cup \{\ell^*\}$\;
    Permanently neutralize residual contribution of block $\ell^*$\;
}

\Return $\mathcal{F}$\;}
\end{algorithm}

}

\subsection{Differential Fault Localization}
\label{subsec:diff-FL}

\newedits{When a clean reference model is available, fault localization can be formulated as a structural comparison problem between the clean and corrupted networks. We refer to this setting as \emph{differential fault localization}. The clean model serves as an oracle representing correct behavior, and faults are identified by detecting structural deviations in intermediate representations as perturbations propagate through the transformer stack.

Our approach, summarized in Algorithm~\ref{alg:bitflipscope}, performs localization in three progressively finer stages: (i) block-level localization, (ii) layer-level localization, and (iii) weight/bit-level localization. Unlike prior heuristic deviation-based approaches, our block localization stage formulates the problem as adaptive change-point detection in the divergence trajectory between clean and faulty hidden states.}

\begin{figure}[h]
    \centering
    \begin{subfigure}{0.4\textwidth}
        \centering
        \includegraphics[width=\linewidth]{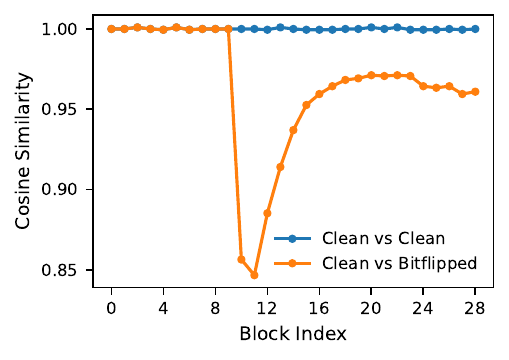}
    \end{subfigure}
    \hfill
    \begin{subfigure}{0.4\textwidth}
        \centering
        \includegraphics[width=\linewidth]{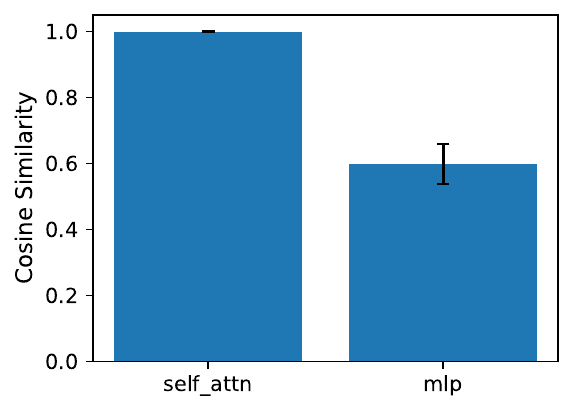}
    \end{subfigure}
    \caption{\newedits{Hidden-state comparison between clean and bit-flipped models for block-level and layer-level localization. (a) Block-wise cosine similarity showing a sharp drop at the corrupted block 9. (b) Cosine similarity across attention and MLP sublayers within block 9. }}
    \label{fig:sim-localization}
\end{figure}

\begin{algorithm}[h]
\caption{BitFlipScope: Differential Fault Localization}
\label{alg:bitflipscope}
\KwIn{Clean model $M_{\text{clean}}$, Faulty model $M_{\text{faulty}}$, Input set $\mathcal{X}$}
\KwOut{Faulty block $B^*$, layer $L^*$, bit indices $\beta^*$}

\BlankLine
\newedits{\textbf{Stage 1: Block Localization (ADCD)}\\
$B^* \leftarrow \texttt{ADCD}(M_{\text{clean}}, M_{\text{faulty}}, \mathcal{X})$}

\BlankLine
\textbf{Stage 2: Layer Localization}\\
$L^* \leftarrow \texttt{LocalizeLayer}(M_{\text{clean}}, M_{\text{faulty}}, B^*, \mathcal{X})$

\BlankLine
\textbf{Stage 3: Weight and Bit Localization}\\
$\beta^* \leftarrow \texttt{LocalizeWeightAndBit}(M_{\text{clean}}, M_{\text{faulty}}, B^*, L^*)$

\BlankLine
\Return $(B^*, L^*, \beta^*)$
\end{algorithm}

\subsubsection{\newedits{Stage 1: Transformer Block Localization via ADCD}}
\newedits{
Let $h_\ell^{\text{clean}}$ and $h_\ell^{\text{faulty}}$ denote hidden states at block $\ell$. 
We compute cosine similarity

\begin{equation}
S_\ell = \cos(h_\ell^{\text{clean}}, h_\ell^{\text{faulty}})
\end{equation}

and define the divergence trajectory

\begin{equation}
d_\ell = 1 - S_\ell.
\end{equation}

Under residual propagation, injected perturbations induce slope discontinuities in $d_\ell$ (see Appendix~\ref{appendix:theory}). 
Fault localization therefore reduces to recovering the support of injected perturbations via change-point detection in this one-dimensional divergence signal \cite{killick2012changepoint}. Fig.~\ref{fig:sim-localization} illustrates how a bit-flip produces a sharp drop in cosine similarity at the corrupted block and its affected sublayer.

\paragraph{Empirical Null Calibration}
Because numerical precision yields small nonzero divergence under clean--clean comparisons, We estimate an empirical null distribution from clean--clean comparisons and define a nonparametric threshold $\tau_{\text{null}}$ as the $99.9$th percentile of this distribution.

\paragraph{Adaptive Deviation Change-Point Detection}
The first faulty block is identified as the earliest layer where $d_\ell > \tau_{\text{null}}$. 
Additional faults are detected by identifying statistically significant positive local maxima in $\Delta_\ell$ using a robust MAD-based scale estimator. The complete procedure is shown in Algorithm~\ref{alg:adcd}.}

\begin{algorithm}[h]
\newedits{
\caption{ADCD Block Localization}
\label{alg:adcd}
\KwIn{Divergence profile $d[0 \dots L-1]$, threshold $\tau_{\text{null}}$}
\KwOut{Faulty block indices $\mathcal{B}$}

Find $\ell_1 = \min\{\ell \mid d_\ell > \tau_{\text{null}}\}$ \;

\If{$\ell_1$ does not exist}{
    \Return $\emptyset$
}

$\mathcal{B} \gets \{\ell_1 - 1\}$ \;

Compute $\Delta_\ell = d_\ell - d_{\ell-1}$ for $\ell > \ell_1$ \;

$m \gets \text{median}(\Delta_\ell)$ \;
$\theta \gets m + 5\,\text{MAD}(\Delta_\ell)$ \;

\For{$\ell = \ell_1+2$ to $L-2$}{
    \If{$\Delta_\ell > \theta$ and local maximum}{
        $\mathcal{B} \gets \mathcal{B} \cup \{\ell - 1\}$
    }
}

\Return $\mathcal{B}$}
\end{algorithm}

\begin{figure}[t]
    \centering
    \includegraphics[width=0.9\columnwidth]{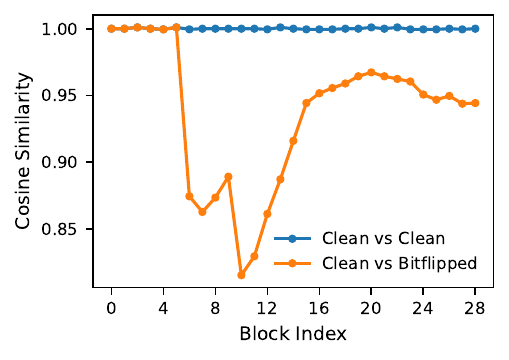}
    \caption{\newedits{Perturbed multiblock(5,9) detection using ADCD in Llama 3.2 3B model showing significant change in cosine similarity at the attacked blocks. Block-level and layer-level similarity analysis is shown in Appendix \ref{appendix:diffref-3b-quant}.}}
    \label{fig:multi_block_analysis_diff}
\end{figure}

\newedits{For multiple‑fault scenarios, BitFlipScope does not stop after identifying the single highest‑scoring block. Instead, we compute robust z‑scores for all blocks and flag every block with z $> \tau$ as potentially corrupted. This global multi‑peak detection allows us to recover multiple attacked blocks in a single run, rather than masking and re‑running the localization procedure. In practice (Fig.~\ref{fig:multi_block_analysis_diff}), bit flips injected into two distinct blocks produce two clear outlier peaks in the z‑score spectrum, both of which are detected without increasing the false positive rate on clean blocks.}

\newedits{
\subsubsection{Stage 2: Layer Localization}
Given suspect blocks $\mathcal{B}$, we compare internal activations of their attention and MLP sublayers across $\mathcal{X}$. The sublayer exhibiting maximal divergence between clean and faulty activations is selected as the faulty layer $L^*$. Because all other blocks remain unperturbed, divergence is spatially localized within the corrupted block, enabling precise identification without exhaustive parameter search.

\subsubsection{Stage 3: Weight and Bit Localization}

Within layer $L^*$, corrupted parameters are isolated via a two-step narrowing procedure.

\paragraph{Parameter Hashing}
Each tensor is hashed (e.g., SHA-1) to rapidly identify mismatches.

\paragraph{Element-wise Comparison}
Only mismatched tensors are compared element-wise to recover the corrupted weights and flipped bit indices $\beta^*$.

This hierarchical narrowing reduces the search from billions of parameters to a constant number of tensors and ultimately individual bits.

\subsubsection{Complexity}
The overall procedure runs in $O(L)$ time with respect to the number of transformer blocks and requires no retraining, gradient computation, or combinatorial search.}

\subsection{Performance Recovery}
\label{subsec:recovery}

Once the corrupted block has been identified, BitFlipScope applies lightweight performance recovery mechanisms to mitigate the impact of the fault without requiring fine-tuning or retraining. An overview of the recovery stage is shown in Fig.~\ref{fig:overview}(c). The recovery strategy differs slightly depending on whether a clean reference model is available.

\paragraph{Self-Referential Setting}
\label{para:performrec_self}
When no clean reference model exists, we cannot directly restore corrupted parameters. Instead, we employ inference-time adjustment mechanisms that reduce the influence of the faulty block on the model’s forward computation.\newedits{Large language models exhibit significant block-level redundancy, where neighbouring transformer blocks generate highly similar representations such that removing selected blocks results in minimal changes to perplexity and overall accuracy \cite{song2024sleb}.} Specifically, we leverage the same residual-scaling mechanism used for localization: by attenuating the corrupted block’s residual contribution, the model’s predictions can be stabilized and partially corrected. This approach selectively suppresses the block’s abnormal computation while leaving the remaining network unchanged, providing a practical and low-cost mitigation method suitable for real-world deployments where reference models or retraining resources may be unavailable.

\paragraph{Differential Setting}
When a clean model is available, BitFlipScope can undertake more targeted recovery. After the corrupted block and faulty sublayer are identified, the parameters of the clean model serve as a ground-truth reference that allows us to restore the affected tensors directly. This correction is performed at the granularity indicated by the localization stage, replacing only the corrupted parameters while preserving all other weights. This targeted restoration avoids the need for global retraining, restores the model’s functional behavior, and ensures minimal disruption to the remaining computation graph.

% \paragraph{Summary}
% In both settings, BitFlipScope enables practical performance recovery by leveraging the fault localization results. The self-referential setting prioritizes inference-time mitigation, while the differential setting supports direct parameter repair. These complementary strategies provide a flexible and lightweight approach for restoring model reliability in the presence of bit-flip faults, as illustrated in Fig.~\ref{fig:overview}(c).

\section{Experiments}

We evaluate BitFlipScope under realistic hardware-fault conditions using a state-of-the-art transformer model subjected to adversarial bit-flip perturbations. Our experiments are designed to assess (1) the accuracy of fault localization under both differential and self-referential settings, and (2) the reliability of the behavioral signals used for diagnosis.

\subsection{Model and Hardware Configuration}

All experiments use the \textbf{LLaMA~3.2 3B}~\cite{grattafiori2024llama3herdmodels} model in 8-bit quantized form. We select this model because it is fully open-source, supports direct weight inspection for controlled bit-flip injection, and provides a realistic yet computationally manageable LLM scale for repeated fault-injection experiments. Its architecture is also representative of larger transformer models~\cite{vaswani2017attention}, making our findings broadly generalizable.

The 8-bit variant is chosen because quantization amplifies the impact of bit-flip faults: with fewer bits per weight, a single flip induces a larger numerical deviation and more pronounced activation distortion, making it an ideal setting for evaluating localization robustness ~\cite{gholami2022survey, Deephammer, GenBFA}. All experiments are conducted on an NVIDIA RTX~6000 Ada GPU (48\,GB).

\subsection{Dataset and Evaluation Metrics}

We use the Massive Multitask Language Understanding (MMLU) benchmark~\cite{hendrycks2021mmlu} to evaluate the model before and after fault injection. MMLU spans a diverse collection of multiple-choice tasks across 57 subject areas, including mathematics, history, law, computer science, and the natural and social sciences. This diversity is well-suited for our study: bit-flip faults may manifest differently depending on the semantic and reasoning demands of the input, and a broad evaluation ensures that we capture fault effects across heterogeneous cognitive workloads rather than a narrow task domain. 

Performance is measured using two metrics:
\begin{itemize}
    \item \textbf{Accuracy}, the fraction of correctly predicted answers across all tasks.
    \item \textbf{Cross-Entropy Loss}, computed as the negative log-likelihood of the correct answer \cite{Fan_Jiang_2025}.
\end{itemize}

Together, these metrics provide both coarse (accuracy) and fine-grained (loss) indicators of degradation, enabling precise assessment of the model’s behavioral changes under bit-flip fault injection.

\subsection{Bit-Flip Injection}

To introduce controlled bit level corruptions, we use the GenBFA methodology~\cite{GenBFA} only to select which weight bits to flip. GenBFA identifies critical parameters using weight magnitude and gradient sensitivity, ensuring that injected faults meaningfully affect model behavior. This selection process is independent of our localization framework, as BitFlipScope does not depend on how flipped bits are chosen.

In our experiments, fewer than 3\% of the candidate bits identified by GenBFA are flipped, producing minimal but high impact perturbations that resemble real hardware faults such as radiation induced single bit errors or targeted fault injection attacks. Each experiment flips one bit at a selected critical location, creating a corrupted model instance used to evaluate localization and recovery.

% \newedits{We adopt GenBFA as our bit‑flip injector in a targeted configuration: the attack selects a small number of highly influential weights and bits to drive the model’s accuracy on MMLU from 67.1\% to 0\%, as in the original work . This configuration is targeted at overall benchmark degradation rather than a single class; thus, faults affect a wide range of prompts and subject areas. BitFlipScope’s self‑referential and differential localization components operate on loss and hidden‑state divergence aggregated over diverse MMLU samples, so they do not assume any specific class distribution and remain agnostic to whether the underlying attack is global or class‑biased.}

\subsection{Evaluation Under Two Reference Settings}

We evaluate BitFlipScope in both deployment scenarios introduced in Section~\ref{sec:methodology}. In the \emph{differential setting}, where a clean reference model is available, localization accuracy is measured by comparing the fault identified by BitFlipScope with the ground-truth corrupted block, sublayer, and parameter. In the \emph{self-referential setting}, where only the corrupted model is available, we assess the ability of residual-scaling-based sensitivity analysis to isolate the faulty block without access to clean activations. Both evaluations are performed across multiple model sizes and corruption sites.

\subsection{Experimental Goals}

Our experiments aim to answer the following questions:

\begin{itemize}
    \item \textbf{Localization Accuracy:} Can BitFlipScope consistently identify the corrupted block or sublayer following a targeted bit-flip?
    \item \textbf{Signal Quality:} Do hidden-state divergences and residual-scaling sensitivities provide clear and separable diagnostic signatures?
    \item \textbf{Robustness Across Models and Fault Sites:} Are localization signals stable across different model sizes, block depths, and corruption locations?
    \item \textbf{Recovery Readiness:} Does accurate localization enable lightweight, non-finetuned performance recovery in both differential and self-referential settings?
\end{itemize}
\begin{figure*}[t]
    \centering

    % Row 1
    \begin{subfigure}{0.48\textwidth}
        \centering
        \includegraphics[width=\linewidth]{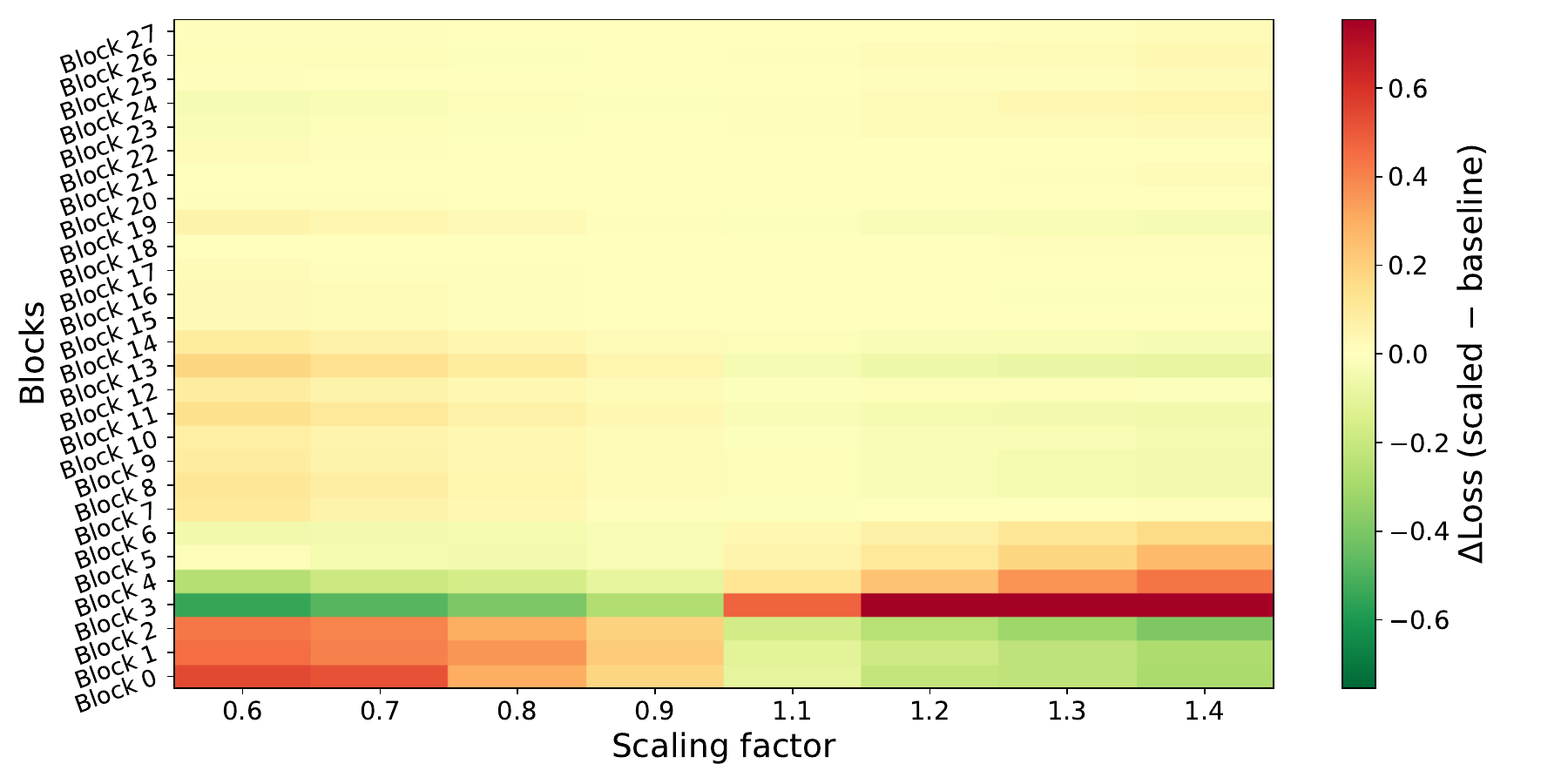}
        \caption{Fault at Block 3}
    \end{subfigure}
    \hfill
    \begin{subfigure}{0.48\textwidth}
        \centering
        \includegraphics[width=\linewidth]{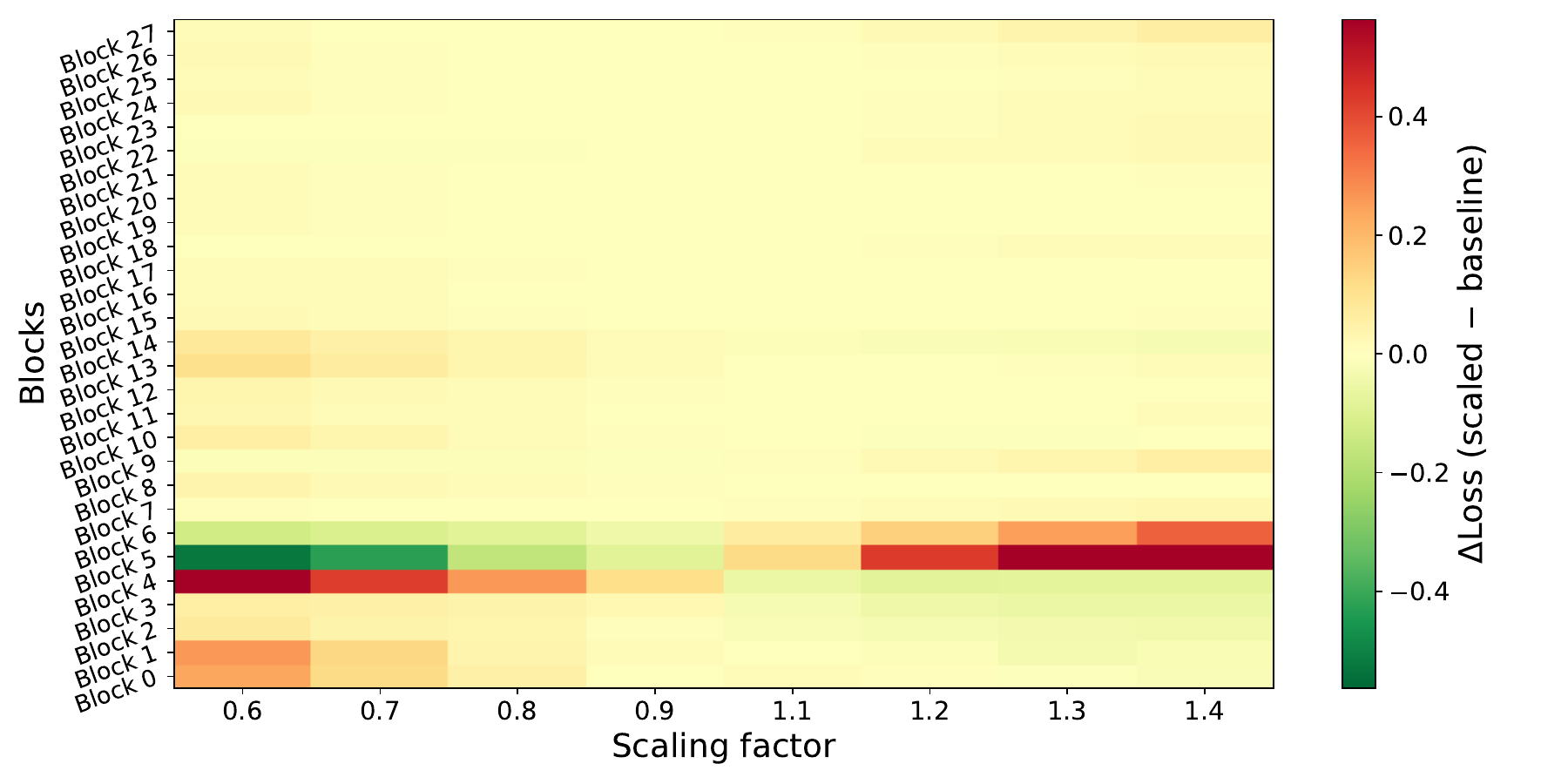}
        \caption{Fault at Block 5}
    \end{subfigure}

    % Row 2
    \begin{subfigure}{0.48\textwidth}
        \centering
        \includegraphics[width=\linewidth]{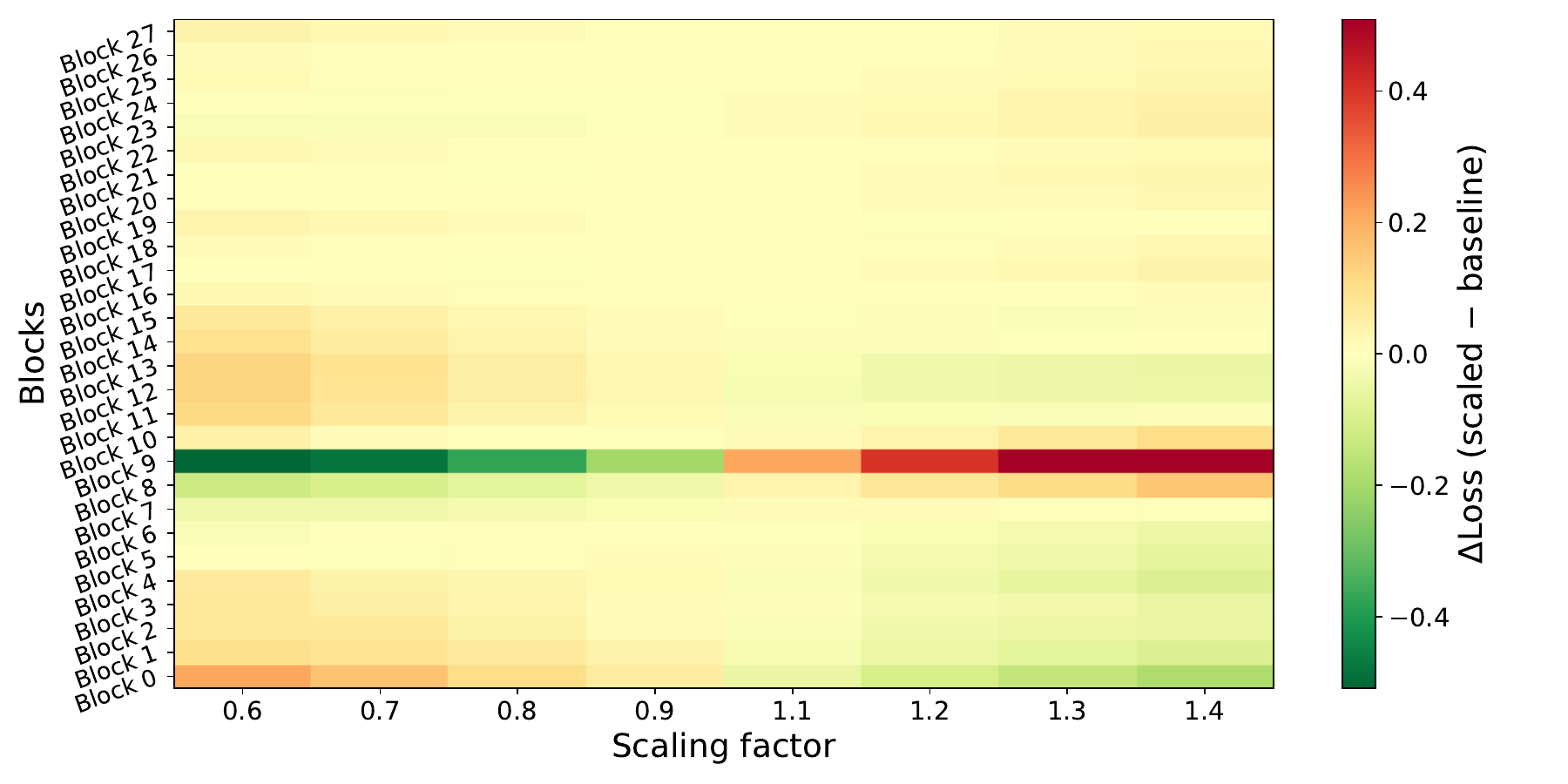}
        \caption{Fault at Block 9}
    \end{subfigure}
    \hfill
    \begin{subfigure}{0.48\textwidth}
        \centering
        \includegraphics[width=\linewidth]{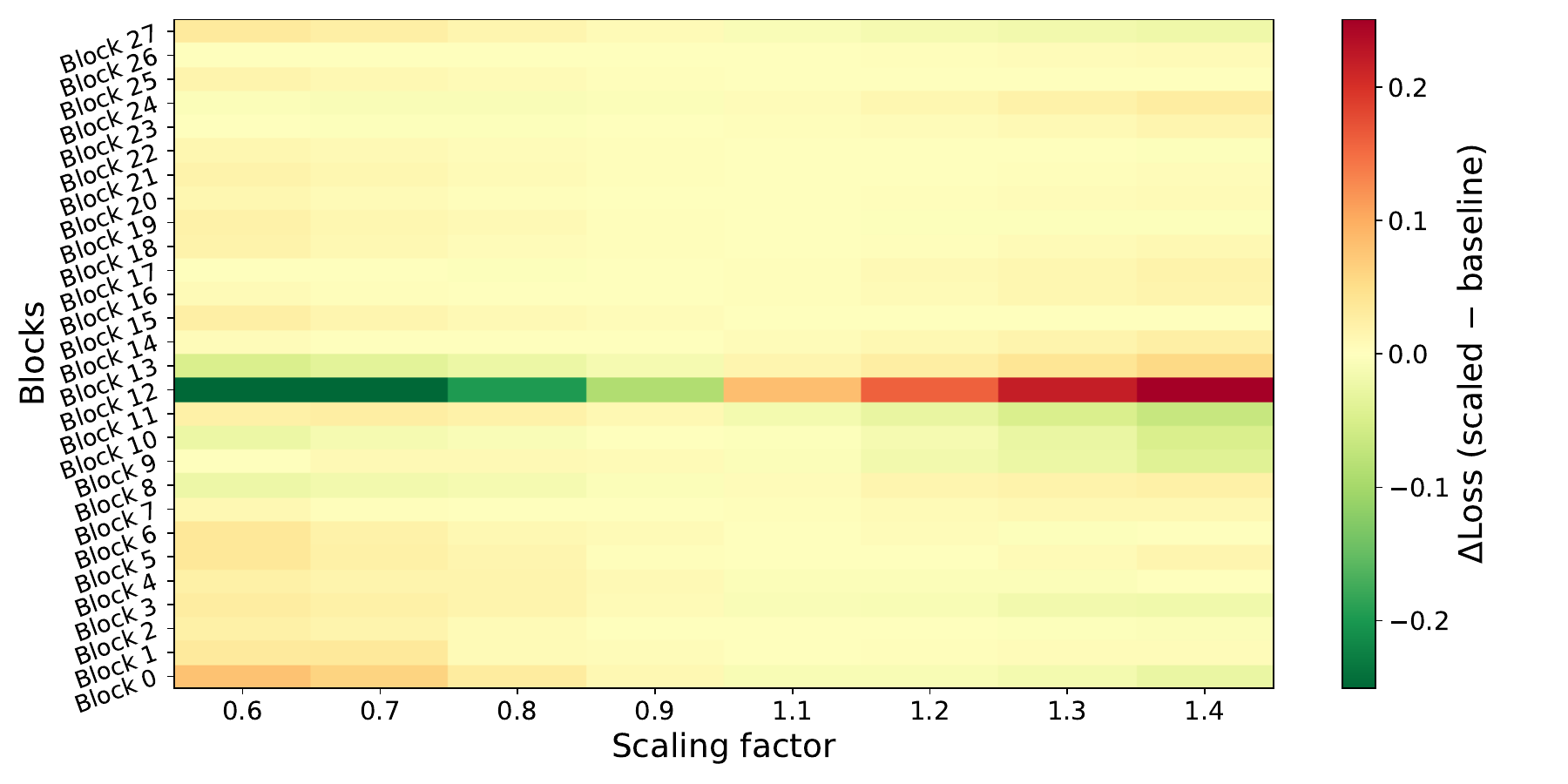}
        \caption{Fault at Block 12}
    \end{subfigure}

    \caption{Heatmaps of $\Delta\mathrm{Loss}$ for LLaMA~3.2 3B across the four injected faults. The corrupted block in each case shows a pronounced asymmetric loss pattern under scaling.}
    \label{fig:selfref-3b-loss}
\end{figure*}
\begin{figure*}[t]
    \centering
    \includegraphics[width=\textwidth]{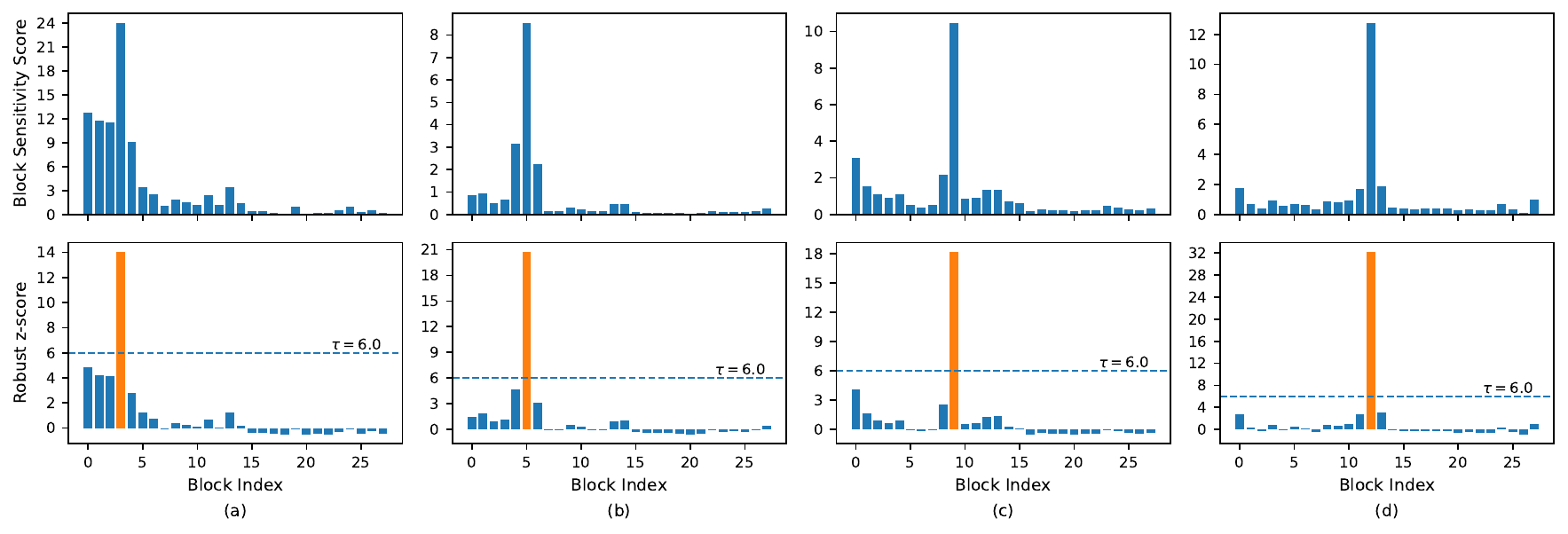}
    \caption{\newedits{Identification of perturbed blocks in LLaMA~3.2 3B under self-referential configurations, using a threshold based on the BSS and robust z-score computed for each block}}
    \label{fig:results}
\end{figure*}

\section{Results}
\label{sec:results}

We evaluate BitFlipScope under both localization settings described in Section~\ref{sec:methodology}. Our results examine (i) how reliably the self-referential approach identifies corrupted blocks using loss sensitivity under residual scaling, (ii) how accurately the differential approach localizes faults at the block and sublayer levels, and (iii) how the computational cost compares to brute-force search. We present results for the two settings separately, followed by an analysis of recovery behavior and overall efficiency.
\begin{figure}[h]
    \centering
    \includegraphics[width=\columnwidth]{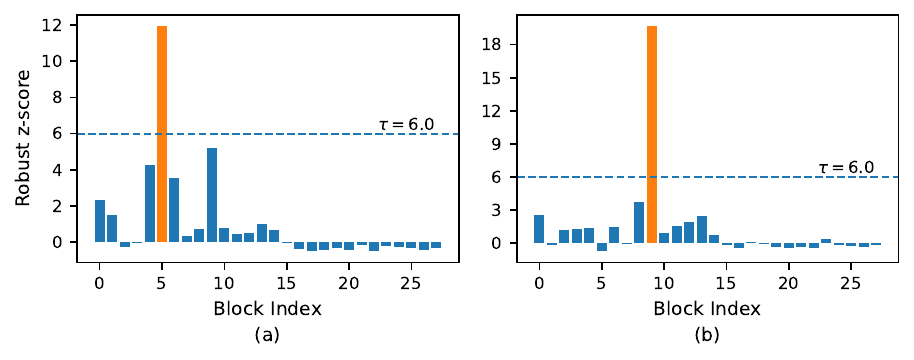}
    \caption{\newedits{Multiple attacked block (5,9) detection in self-referential settings of Llama 3.2 3B model leveraging the threshold}}
    \label{fig:multi_block_analysis_5_9}
\end{figure}

\begin{figure}[t]
    \centering
    \includegraphics[width=\columnwidth]{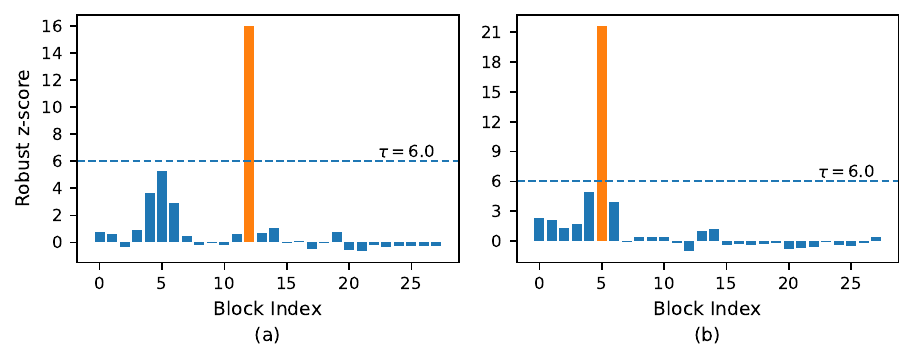}
    \caption{\newedits{Multiple attacked block (12,5) detection in self-referential settings of Llama 3.2 3B model leveraging the threshold}}
    \label{fig:multi_block_analysis_12_5}
\end{figure}
% 1. Short section intro
\subsection{Self-Referential Fault Localization Results} 
\label{subsec: self-ref_FL_results}
We first evaluate the self-referential localization method on the LLaMA~3.2 3B model by injecting bit-flips into four critical blocks identified by GenBFA~\cite{GenBFA} (blocks 3, 5, 9, and 12). Fig.~\ref{fig:selfref-3b-loss} shows in each case, the corrupted block exhibits a distinctive asymmetric pattern: the loss sharply increases for $\alpha > 1$ and decreases for $\alpha < 1$, whereas non-faulty blocks show minimal or monotonic behavior. This strong contrast provides a clear signal for isolating the faulty block.

\newedits{For residual‑scaling based localization, we evaluate BitFlipScope using a validation subset of 256 MMLU examples per run. Empirically, we find that BSS ranking of blocks converges rapidly: with as few as 64 examples, the corrupted block attains the maximum BSS in all four tested injection sites, and increasing the batch to 256 or 512 samples further reduces variance but does not change the top‑ranked block. This suggests that reliable block identification can be achieved with relatively modest batches, keeping the overall cost dominated by a small number of forward passes rather than large‑scale dataset sweeps.}

\newedits{To quantify this behavior, we compute the Block Sensitivity Score (BSS) for each block and leverage the robust sensitivity-based detection that computes a threshold, $\tau$ = 6.0 with a robust z-score which effectively detects the perturbed blocks(3, 5, 9 \& 12) shown in Fig.~\ref{fig:results}. For our experiments, we sweep $\alpha$ from 0.6 to 1.4 in steps of 0.1, leading to 256 forward passes for 32 blocks. However, the heatmaps in Fig.~\ref{fig:selfref-3b-loss} clearly indicate that if we restrict the sweep to either 0.6 and 1.4 or 0.7 and 1.3, only 64 forward passes are needed. For a trillion-parameter model such as GLaM \cite{glam}, which has 64–96 blocks, this corresponds to 132–192 forward passes, which is not computationally expensive.}

\newedits{While later blocks (e.g., Block 12 in Fig.~\ref{fig:selfref-3b-loss}) exhibit smaller absolute $\Delta$Loss magnitudes than early blocks under the same $\alpha $‑sweep, BitFlipScope operates on normalized sensitivity scores rather than raw loss differences. In each injection, we compute per‑block BSS and then standardize scores using a robust normalization described in Section~\ref{sec:methodology}. This normalization removes global depth trends and ensures that even later‑depth faults with smaller absolute $\Delta$Loss still appear as statistically significant outliers relative to intact blocks. In all four injected faults, the corrupted block achieves the maximum robust z‑score, irrespective of depth.}

To assess scalability, we repeat the same procedure on the larger LLaMA~3.1 8B model, injecting bit-flips into blocks 3, 9, 15, and 28. BitFlipScope again identifies the correct corrupted block in every case, exhibiting the same asymmetric $\Delta\mathrm{Loss}$ pattern and dominant BSS score observed in the 3B model. \newedits{Heatmaps, sensitivity and robust z-score plots computed from each block, with threshold-based detection for the Llama 3.1 8B model, are provided in Appendix~\ref{appendix:results_8b}. Optimal selection of the threshold and avoiding any false positive detection results are provided in Appendix \ref{appendix:robust_detection} \& \ref{appendix:selfref-noFP}.}These results demonstrate that residual-path sensitivity generalizes consistently to deeper and larger LLM architectures.\\
\newedits{
We next evaluate BitFlipScope in scenarios where multiple transformer blocks are corrupted simultaneously. Figs.~\ref{fig:multi_block_analysis_5_9} and~\ref{fig:multi_block_analysis_12_5} show representative cases where bit-flips are injected into two layers of the LLaMA-3.2 3B model. In both settings, the robust z-score based detection identifies the dominant corrupted block in the first iteration and subsequently reveals the second corrupted block once the first fault is neutralized. These results confirm that BitFlipScope can reliably localize multiple corrupted blocks within the same model.
}\\
\newedits{
Finally, we evaluate false positives by applying BitFlipScope’s self-referential pipeline to a clean LLaMA-3.2 3B model under the same $\alpha$-sweep. In this setting, no block exceeds the corruption threshold $\tau = 3$ across $N$ independent runs, yielding a false positive rate of 0\%. The corresponding BSS and robust z-score distributions for the clean model are shown in Appendix~\ref{appendix:selfref-noFP}.}
\subsection{Differential Fault Localization Results}
We evaluate BitFlipScope’s differential localization performance by injecting single bit-flips into both MLP and self-attention sublayers across multiple transformer blocks and model sizes. In this setting, the clean model serves as a behavioral reference, and localization is determined by identifying block and sublayer divergences between the clean and corrupted models. Table~\ref{tab:merged-results} summarizes the outcomes for the tested bit-flip locations.

% ------------------------- MLP TABLE --------------------------
\begin{table}[h]
\centering
\caption{Localization success for bit-flips injected into MLP (up/down) and Attention (Q/K/V) sublayers across model blocks.}
\small
\renewcommand{\arraystretch}{1.1}
\begin{tabular}{lcccc}
\toprule
\textbf{Model} & \textbf{Block} & \makecell{\textbf{MLP} \\ \textbf{(up/down)}} & \makecell{\textbf{Attention} \\ \textbf{(Q/K/V)}} & \textbf{Localized} \\
\midrule
LLaMA 3B & 3 & up, down & q, k, v & $\checkmark$ \\
LLaMA 3B & 5 & up, down & q, k, v & $\checkmark$ \\
LLaMA 3B & 9 & up, down & q, k, v & $\checkmark$ \\
LLaMA 3B & 12 & up, down & q, k, v & $\checkmark$ \\
LLaMA 8B & 3 & up, down & q, k, v & $\checkmark$ \\
LLaMA 8B & 9 & up, down & q, k, v & $\checkmark$ \\
LLaMA 8B & 15 & up, down & q, k, v & $\checkmark$ \\
LLaMA 8B & 28 & up, down & q, k, v & $\checkmark$ \\
\bottomrule
\end{tabular}
\label{tab:merged-results}
\end{table}

\paragraph{Sublayer Localization}
BitFlipScope reliably localizes faults within both MLP and attention sublayers across all tested blocks and model sizes. Faults injected into MLP up- and down-projection layers as well as attention query, key, and value projections are consistently isolated through differential hidden-state comparison. Detailed similarity measurements are provided in Appendix~\ref{appendix:diffref-3b-quant}.

\paragraph{Generalization Across Models and Blocks}
The consistent localization performance across a wide range of blocks and across two model sizes (3B \& 8B) indicates that the differential method generalizes well with respect to architectural depth, scale, and parameter type.

\subsubsection{Efficiency Compared to Brute Force}
Exhaustive parameter differencing requires comparing all model parameters. For a representative 16-block transformer this results in $\approx1.88\times10^9$ elementwise comparisons, whereas BitFlipScope requires $\approx1.68\times10^7$ comparisons after hierarchical filtering. Details of this calculation are provided in Appendix \ref{appendix:bf-comparison}.

\subsection{Performance Recovery Results}
After localizing the corrupted block, we evaluate whether BitFlipScope enables practical, fine-tuning-free recovery in the self-referential setting. We mitigate the bit-flip fault by zeroing out the residual contribution of the identified faulty block, thereby suppressing its corrupted computation. Table~\ref{tab:recovery-summary} summarizes the resulting MMLU accuracy.

For the LLaMA 3.2 3B model, accuracy drops from 61\% to 3.2\% after a bit-flip but recovers to 51\% after mitigation, restoring 82.7\% of the lost performance. Similarly, for the LLaMA 3.1 8B model, accuracy recovers from 3.9\% back to 56\%, recovering 80.0\% of the degradation. These results demonstrate that once BitFlipScope identifies the corrupted block, inexpensive inference-time mitigation can substantially restore model behavior without retraining or a clean reference model.

In the differential setting, where a clean reference model is available, BitFlipScope enables full restoration of model functionality: replacing the corrupted weight tensor with its clean counterpart fully recovers baseline accuracy in every evaluated case.

% \newedits{For LLaMA‑3.2 3B, a full tensor‑by‑tensor parameter diff requires reading and comparing approximately 6 GB of quantized weights for each differential check. In contrast, BitFlipScope’s staged differential pipeline first performs block‑wise and layer‑wise activation comparisons (dominated by a few forward passes), and then hashes and compares only tensors within the localized layer, which in our experiments corresponded to less than 0.5\% of the total parameters. This reduces the effective comparison volume by more than two orders of magnitude while achieving the same bit‑level localization granularity.}

% \newedits{While our experiments use GenBFA to select impactful bit locations in quantized LLaMA models, BitFlipScope is agnostic to the mechanism that chooses which bits are flipped. For attacks such as PBS that operate in a different architecture and training regime, the primary effect—localized weight corruption yielding abnormal hidden‑state divergence and loss sensitivity—remains the same. Our differential setting in particular would continue to localize faults at block, layer, and bit levels as long as a clean reference is available. Porting PBS directly to the LLaMA‑3.x family is non‑trivial and outside the scope of this work, but we expect BitFlipScope to remain applicable because it only assumes bit‑induced deviations in internal activations, not a specific attack heuristic.}

\begin{table}[htbp]
\centering
\caption{MMLU accuracy before attack, after bit-flip, and after recovery.}
\small
\begin{tabular}{lcccc}
\toprule
\textbf{Model} &
\makecell{\textbf{Baseline} \\ \textbf{Accuracy}} &
\makecell{\textbf{After} \\ \textbf{Bitflip}} &
\makecell{\textbf{After} \\ \textbf{Recovery}} &
\makecell{\textbf{Recovery} \\ \textbf{Percentage}} \\
\midrule
LLaMA 3B & 61.0\% & 3.2\% & 51.0\% & 82.7\% \\
LLaMA 8B & 69.0\% & 3.9\% & 56.0\% & 80.0\% \\
\bottomrule
\end{tabular}
\label{tab:recovery-summary}
\end{table}

\section{Conclusion}

This paper introduced \textbf{BitFlipScope}, a unified and scalable framework for localizing and mitigating bit-flip faults in large language models under two practical deployment settings: with and without access to a clean reference model. By leveraging differential behavioral analysis in the former case and residual-path sensitivity in the latter, BitFlipScope provides reliable and interpretable diagnostic signals that accurately localize corrupted blocks and sublayers while requiring orders of magnitude fewer computations than brute-force search. Our findings further show that identified faults can be effectively mitigated: targeted parameter replacement fully restores accuracy when a clean model is available, and inference-time attenuation recovers over 80\% of lost performance in the self-referential setting without fine-tuning. These results demonstrate that scalable, software-only fault localization is both feasible and impactful for modern multi-billion-parameter models, laying the groundwork for more robust, secure, and self-diagnosing LLM deployments in safety-critical environments.

\clearpage
\bibliographystyle{IEEEtran}
\bibliography{BitFlipScope}

\clearpage
\appendices
\section{Selection of Residual Scaling Values}
\label{appendix:alpha_sweep}

To determine which residual scaling values provide the most informative diagnostic signal, we perform an empirical $\alpha$-sweep. For a representative block, the residual scaling parameter is varied across a broad range $\alpha \in [0.2,1.8]$, and the corresponding loss change $\Delta\mathrm{Loss}(\ell,\alpha)$ is measured.

\begin{figure}[htbp]
    \centering
    \includegraphics[width=\linewidth]{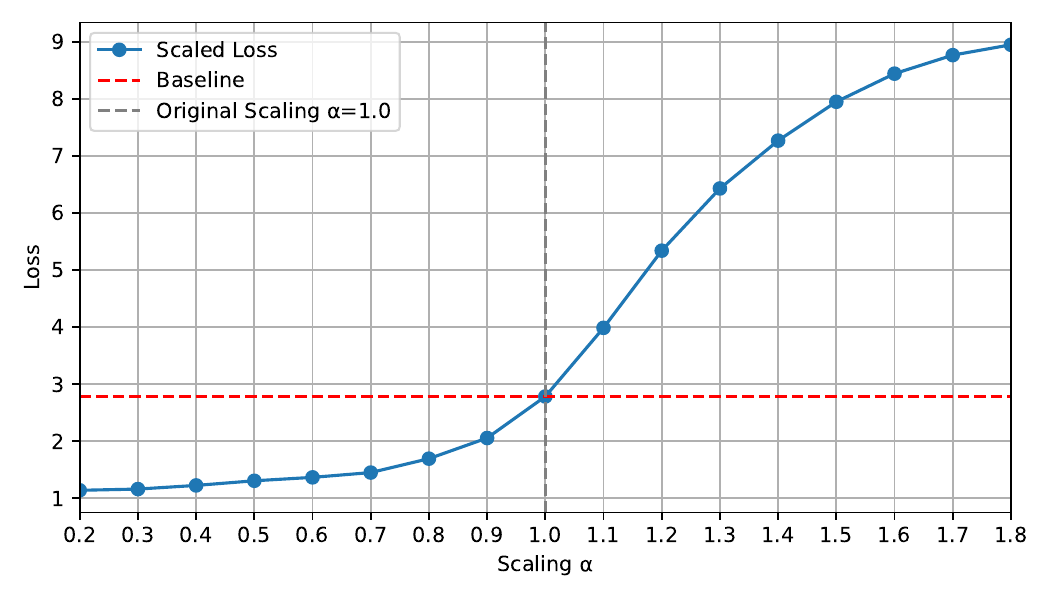}
    \caption{Loss change $\Delta\mathrm{Loss}$ across a broad range of scaling values. Diagnostic sensitivity is highest in the interval $[0.6,1.4]$, motivating the scaling values used in the experiments.}
    \label{fig:alpha-sweep}
\end{figure}

As illustrated in Fig.~\ref{fig:alpha-sweep}, loss changes are most pronounced near moderate perturbations of the residual pathway but flatten for extreme scaling values. When $\alpha < 0.6$, the residual contribution is heavily suppressed, saturating the effect of the block and producing limited diagnostic variation. Conversely, when $\alpha > 1.4$, amplification of the residual pathway causes the block output to dominate the hidden state, and further increases produce diminishing changes in loss.

Based on this empirical observation, we restrict the scaling values used in our experiments to the interval $[0.6,1.4]$. In practice, we use the discrete set
\[
\alpha \in \{0.6,\,0.7,\,0.8,\,0.9,\,1.1,\,1.2,\,1.3,\,1.4\},
\]
which provides a balanced exploration of attenuation and amplification around the nominal scaling $\alpha=1$.

\section{Theoretical Foundations of Adaptive Divergence Change-Point Detection}
\label{appendix:theory}
\subsubsection{Residual Error Propagation Model}

We formalize the behavior of hidden-state divergence under fault injection.

Consider a transformer with residual blocks:

\begin{equation}
h_\ell = h_{\ell-1} + F_\ell(h_{\ell-1}).
\end{equation}

Let the clean and faulty hidden states be $h_\ell^{c}$ and $h_\ell^{f}$, respectively, and define the deviation:

\begin{equation}
\delta_\ell = h_\ell^{f} - h_\ell^{c}.
\end{equation}

If a fault is injected at block $k$, we model it as an additive perturbation $\epsilon_k$:

\begin{equation}
\delta_{k+1} = \delta_k + \epsilon_k.
\end{equation}

For subsequent layers $\ell > k$, propagation obeys:

\begin{equation}
\delta_{\ell+1} = \delta_\ell + J_\ell \delta_\ell,
\end{equation}

where $J_\ell$ is the local Jacobian of $F_\ell$.

Assuming $F_\ell$ is locally Lipschitz with constant $L_\ell$:

\begin{equation}
\|\delta_{\ell+1}\| \le (1 + L_\ell)\|\delta_\ell\|.
\end{equation}

Thus, injected perturbations propagate smoothly after their point of introduction.

\subsection{Structure of the Divergence Trajectory}

We measure divergence via cosine similarity:

\begin{equation}
d_\ell = 1 - \cos(h_\ell^{c}, h_\ell^{f}).
\end{equation}

For small deviations and unit-normalized representations, a second-order Taylor expansion of cosine similarity yields
\[
d_\ell \approx \frac{1}{2}\|\delta_\ell\|^2,
\]
up to higher-order terms. Therefore, under sparse fault injection, the divergence trajectory can be approximated as:

\begin{equation}
d_\ell = g_\ell + \sum_{k \in \mathcal{F}} a_k \mathbf{1}_{\ell \ge k},
\end{equation}

where:

\begin{itemize}
\item $g_\ell$ is a smooth function induced by residual propagation,
\item $\mathcal{F}$ is the set of faulty blocks,
\item $a_k$ represents the magnitude of injected perturbation.
\end{itemize}

This representation shows that faults induce discrete slope discontinuities in $d_\ell$.

\subsection{Reduction to Sparse Support Recovery}

Define first differences:

\begin{equation}
\Delta_\ell = d_\ell - d_{\ell-1}.
\end{equation}

Under the above model:

\begin{equation}
\Delta_\ell = a_\ell + \eta_\ell,
\end{equation}

where:

\begin{itemize}
\item $a_\ell$ is nonzero only at faulty blocks,
\item $\eta_\ell$ represents smooth propagation noise.
\end{itemize}

Therefore, identifying faulty blocks reduces to recovering the support of $a_\ell$ in a one-dimensional sparse signal corrupted by bounded noise.

\subsection{Robust Outlier Detection}

we estimate dispersion using Median Absolute Deviation (MAD):

\begin{equation}
\text{MAD} = \text{median}(|\Delta_\ell - \text{median}(\Delta_\ell)|).
\end{equation}

MAD has a breakdown point of 50\% and provides a robust scale estimator without distributional assumptions. Under high signal-to-noise ratio (SNR),

\begin{equation}
\min_{k \in \mathcal{F}} a_k \gg \max_\ell |\eta_\ell|,
\end{equation}

the support of $a_\ell$ is recoverable via thresholding:

\begin{equation}
\Delta_\ell > \text{median}(\Delta_\ell) + 5 \cdot \text{MAD}.
\end{equation}

\paragraph{Threshold Selection.}
The constant factor $5$ in the MAD-based threshold corresponds to approximately $3.4\sigma$ under Gaussian noise, yielding a conservative high-confidence outlier detector. Importantly, because both $\Delta_\ell$ and $\text{MAD}$ scale proportionally with signal magnitude, the rule is dimensionless and invariant to model scale. In practice, injected perturbations exceed propagation ripple by one to two orders of magnitude, making recovery insensitive to moderate variations in this constant.

Because transformer depth induces systematic variation in loss sensitivity, especially between shallow and deep blocks, we do not use a single fixed absolute threshold. Instead, we apply a robust z‑score over the BSS distribution across blocks. Blocks with z‑score greater than $\tau$ (e.g., $\tau$ = 6) are flagged as corrupted, which automatically adapts to depth‑dependent sensitivity patterns without requiring manually tuned, layer‑specific thresholds.
\begin{figure*}[t]
\centering

\begin{subfigure}{\textwidth}
    \centering
    \includegraphics[width=\linewidth]{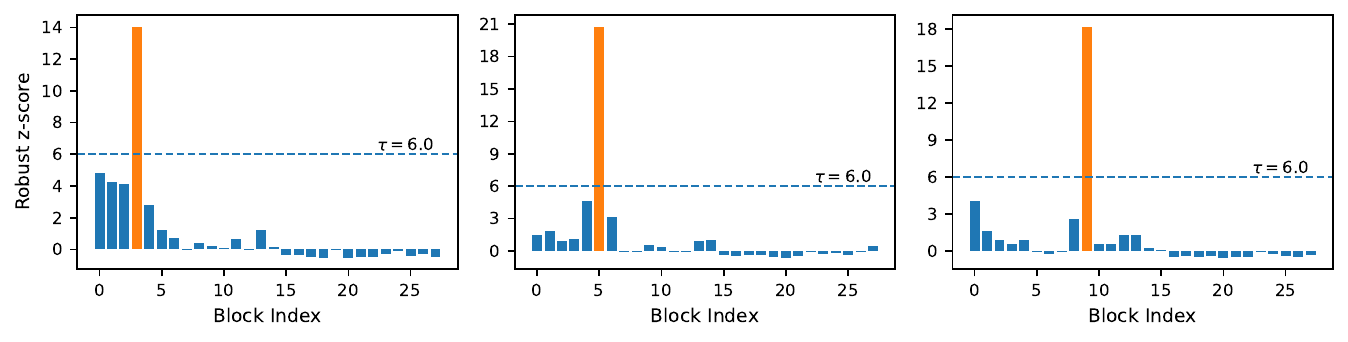}
    \label{fig:block_similarity}
\end{subfigure}

\begin{subfigure}{\textwidth}
    \centering
    \includegraphics[width=\textwidth]{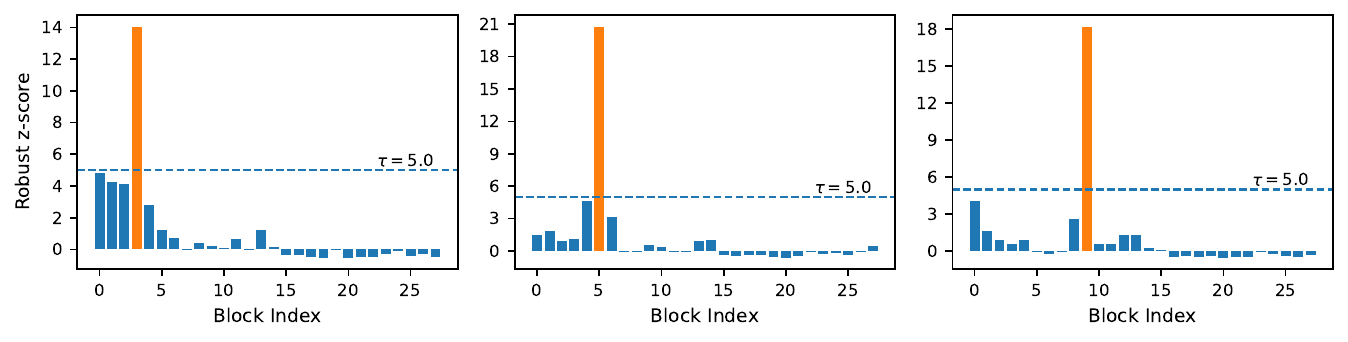}
    \label{fig:layer_similarity}
\end{subfigure}

\begin{subfigure}{\textwidth}
    \centering
    \includegraphics[width=\textwidth]{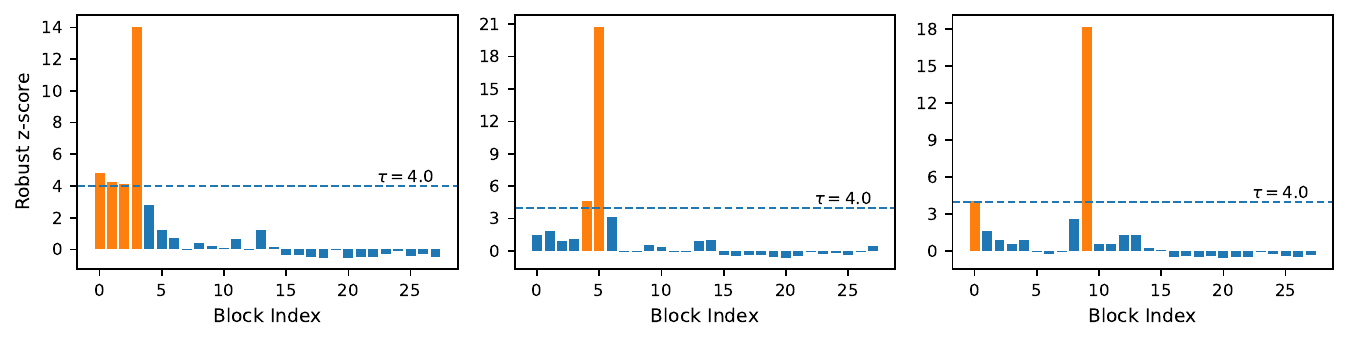}
    \label{fig:projection_comparison}
\end{subfigure}

\caption{Comparison of different threshold values $\tau$ to showcase the optimal setting that minimizes false positives}
\label{fig:full_similarity_analysis}

\end{figure*}
\subsection{Justification of Empirical Null Calibration}

In finite-precision arithmetic, clean--clean comparisons yield small but nonzero divergence due to quantization and numerical effects.

We estimate the empirical null distribution:

\begin{equation}
\mathcal{D}_{\text{null}} = \{\Delta_\ell^{\text{clean-clean}}\}.
\end{equation}

\begin{equation}
\tau_{\text{null}} = \text{Percentile}_{99.9}(\mathcal{D}_{\text{null}})
\end{equation}

ensures controlled false positive rate without assuming Gaussianity. This yields a nonparametric Neyman--Pearson detector with empirical calibration.

\subsection{Sufficient Condition for Exact Recovery}

Let $\eta_\ell$ denote propagation noise and $a_k$ injected magnitudes. 
If

\begin{equation}
\min_{k \in \mathcal{F}} a_k >
\gamma \cdot \sup_\ell |\eta_\ell|,
\end{equation}

for some constant $\gamma > 1$ determined by the MAD thresholding rule, then the support of $a_\ell$ is exactly recoverable.

\section{Statistical Justification of Robust Sensitivity-Based Detection}
\label{appendix:robust_detection}

\subsection{Robust Normalization via Median and MAD}

The Block Sensitivity Scores (BSS) are aggregated across layers to quantify the response of each block to residual perturbation. Since corrupted layers may induce heavy-tailed deviations, we employ a robust normalization scheme based on the median and Median Absolute Deviation (MAD) \cite{LEYS2013764}.

Let $\{\mathrm{BSS}(\ell)\}_{\ell=1}^{L}$ denote the sensitivity scores across $L$ layers. We compute the median
\[
m = \mathrm{median}\big(\mathrm{BSS}(\ell)\big),
\]
and the Median Absolute Deviation
\[
\mathrm{MAD} = \mathrm{median}\big(|\mathrm{BSS}(\ell) - m|\big).
\]
The robust scale estimator is defined as
\[
s = 1.4826 \cdot \mathrm{MAD},
\]
where the constant ensures consistency under Gaussian noise. The resulting robust z-score for layer $\ell$ is
\[
z(\ell) = \frac{\mathrm{BSS}(\ell) - m}{s}.
\]

Both the median and MAD possess a breakdown point of 50\%, meaning that up to half of the layers may exhibit arbitrary deviations without destabilizing the estimator. This makes the normalization procedure well-suited for sparse fault settings.

\subsection{Threshold Selection via Extreme-Value Theory}

Under the null hypothesis that all layers are healthy, the normalized sensitivity scores can be modeled as sub-Gaussian random variables. For a single layer, standard concentration inequalities yield
\[
\mathbb{P}(z(\ell) > t) \leq \exp(-t^2/2).
\]

To control false positives across all $L$ layers, we apply a union bound:
\[
\mathbb{P}\big(\max_{\ell} z(\ell) > t\big)
\leq
L \exp(-t^2/2).
\]

To achieve a family-wise false positive rate bounded by $\alpha$, it suffices to choose
\[
t \geq \sqrt{2 \log\left(\frac{L}{\alpha}\right)}.
\]

Thus, the detection threshold scales as $\mathcal{O}(\sqrt{\log L})$, growing only logarithmically with model depth. In practice, we select $\tau$ slightly above this bound (typically in the range 6–8 for models with 20–100 layers), providing strong false-positive control while maintaining sensitivity to moderate faults.

\subsection{Iterative Detection and Multi-Fault Robustness}

When multiple corrupted layers are present, dominant faults may partially inflate the dispersion of the BSS distribution, thereby masking weaker anomalies. To mitigate such masking effects, we adopt an iterative detection strategy: once a layer exceeds the detection threshold, its residual contribution is neutralized, and the sensitivity scores are recomputed over the remaining layers. 

Because the median and MAD remain stable under sparse contamination, successive iterations progressively expose weaker faults without destabilizing the estimator, enabling reliable identification of subtle anomalies.

This procedure enables reliable localization of multiple corrupted blocks while preserving statistical control over spurious detections, as illustrated in Fig.~\ref{fig:full_similarity_analysis}.

\section{Block-level and layer-level similarity analysis for LLaMA 3.2 3B model}
\label{appendix:diffref-3b-quant}
To further illustrate BitFlipScope’s differential localization behavior, we report block-level and layer-level similarity scores between the clean and corrupted models. Similarity is computed using cosine similarity between hidden-state activations of the corresponding blocks and sublayers. Lower similarity indicates stronger divergence caused by the injected bit-flip.

Table~\ref{tab:block_layer_similarity} summarizes representative cases for the LLaMA 3.2 3B model. The corrupted block consistently exhibits the largest deviation from the clean reference, and the affected sublayer shows the lowest similarity within the block, enabling precise localization.

\begin{table}[h]
\centering
\caption{Block-level and layer-level similarity analysis for the Llama 3.2 3B model}
\small
\begin{tabular}{ccccc}
\toprule
\textbf{Model} & \multicolumn{2}{c}{\textbf{Faulty Block}} & \multicolumn{2}{c}{\textbf{Faulty Layer}} \\
\cmidrule(lr){2-3} \cmidrule(lr){4-5}
 & \textbf{Index} & \textbf{Similarity} & \textbf{Name} & \textbf{Similarity} \\
\midrule
Llama 3.2 3B & 12 & 0.81 & up\_proj   & 0.38 \\
Llama 3.2 3B & 9  & 0.85 & down\_proj & 0.55 \\
Llama 3.2 3B & 5  & 0.84 & down\_proj & 0.59 \\
\bottomrule
\end{tabular}
\label{tab:block_layer_similarity}
\end{table}

\section{Additional Self-Referential Results for LLaMA 3.2 3B with no false positive attacked block detection in terms of clean model}
\label{appendix:selfref-noFP}
To evaluate false positives, we apply the self-referential localization pipeline to a clean LLaMA 3.2 3B model under the same $\alpha$-sweep used in the main experiments. For each block, we compute the Block Sensitivity Score (BSS) and the corresponding robust z-score.

\begin{figure}[h]
    \includegraphics[width=\linewidth]{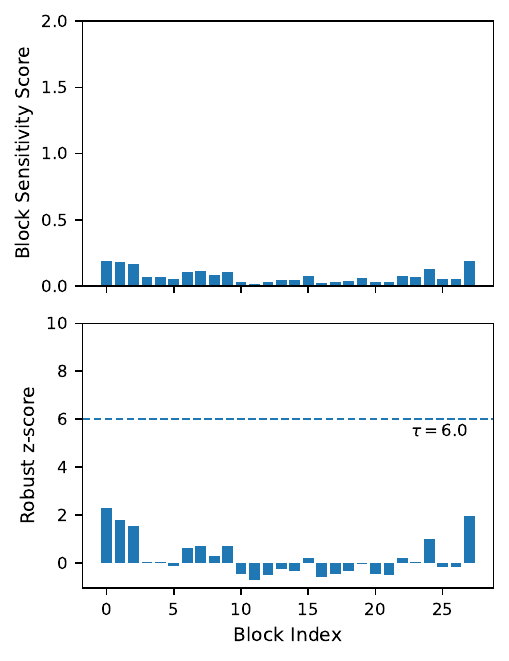}
    \caption{Demonstration of no false positive attacked block detection in clean LLaMA 3.2 3B under self-referential configurations, using a threshold based on the BSS and robust z-score computed for each block}
    \label{fig:block_analysis_FP}
\end{figure}

Fig.~\ref{fig:block_analysis_FP} shows the resulting sensitivity distribution across all blocks. No block exceeds the corruption threshold, confirming that BitFlipScope does not incorrectly flag any block as faulty in the clean model setting.

\section{Comparison with Brute-Force Parameter Differencing}
\label{appendix:bf-comparison}
To contextualize the efficiency of BitFlipScope, we compare it against exhaustive parameter differencing. 
Consider a transformer with 16 blocks, each containing 7 tensors of size $16{,}777{,}216$. 
A brute-force comparison therefore requires

\[
16 \times 7 \times 16{,}777{,}216
= 1{,}879{,}419{,}392
\]

elementwise comparisons.

In contrast, BitFlipScope performs hierarchical filtering:

\begin{itemize}
\item 16 hidden-state comparisons (block filtering),
\item 2 activation comparisons (layer filtering),
\item 3 tensor hash computations,
\item 1 tensor-level comparison of size $16{,}777{,}216$.
\end{itemize}

The resulting number of comparisons is

\[
16 + 2 + 3 + 16{,}777{,}216
= 16{,}777{,}237.
\]

Thus, BitFlipScope reduces the number of comparisons by more than two orders of magnitude relative to brute-force parameter differencing.

\begin{table}[h]
\centering
\caption{Computation cost comparison between brute-force parameter differencing and BitFlipScope.}
\small
\begin{tabular}{lc}
\toprule
\textbf{Method} & \textbf{Number of Comparisons} \\
\midrule
Brute Force   & $1.88 \times 10^{9}$ \\
BitFlipScope  & $1.68 \times 10^{7}$ \\
\bottomrule
\end{tabular}
\label{tab:bf-comparison}
\end{table}

\section{Additional Self-Referential Results for LLaMA 3.1 8B}
\label{appendix:results_8b}
To demonstrate that the self-referential localization behavior generalizes to larger models, we replicate the same analysis on the LLaMA~3.1 8B model. Bit-flips are injected into blocks 3, 9, 15, and 28, and we compute the loss change under the same residual-scaling sweep used in the main experiments.

\begin{figure*}[h]
\centering

\begin{subfigure}{0.48\linewidth}
\centering
\includegraphics[width=\linewidth]{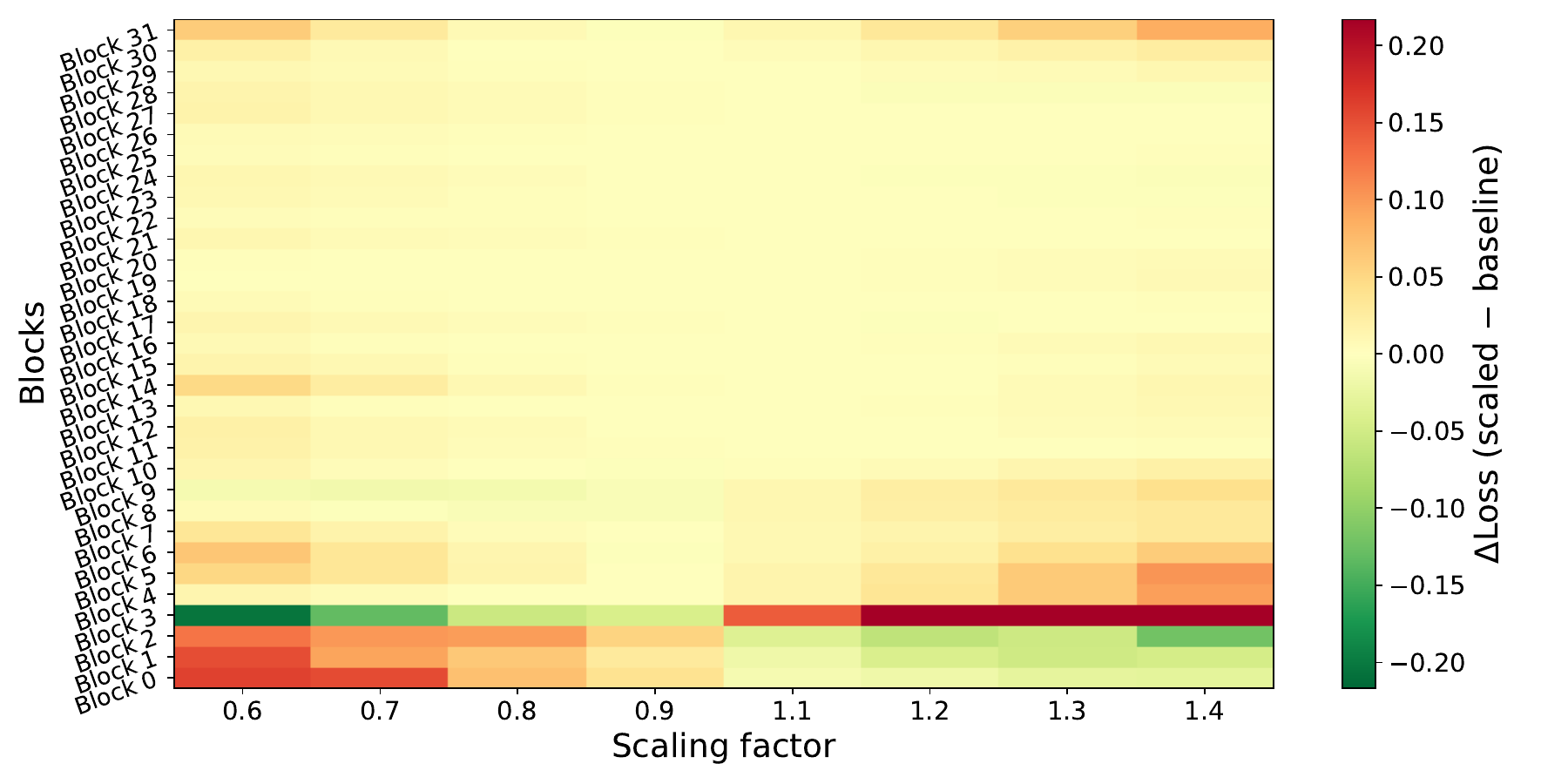}
\caption{}
\end{subfigure}
\hfill
\begin{subfigure}{0.48\linewidth}
\centering
\includegraphics[width=\linewidth]{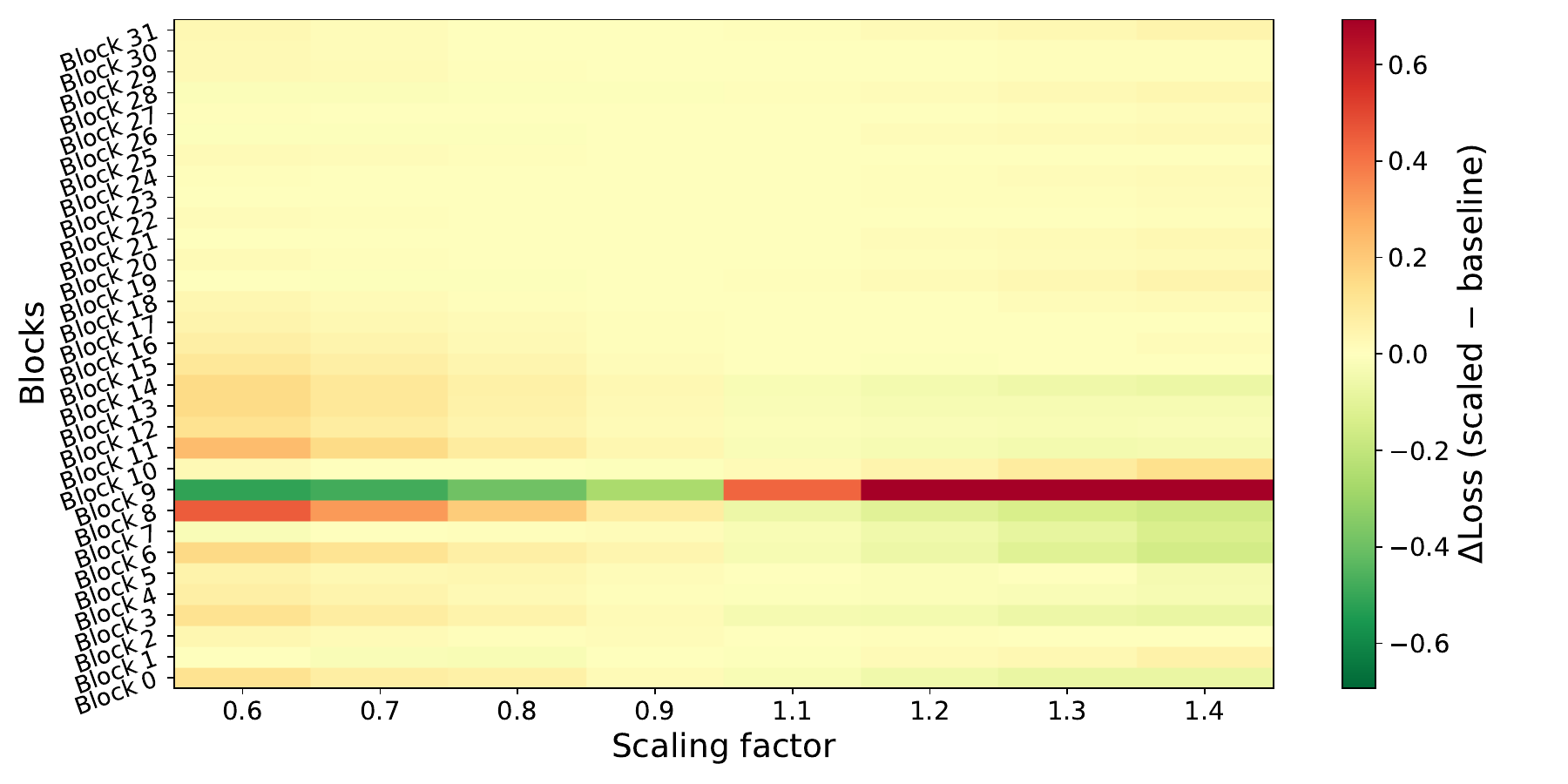}
\caption{}
\end{subfigure}

\begin{subfigure}{0.48\linewidth}
\centering
\includegraphics[width=\linewidth]{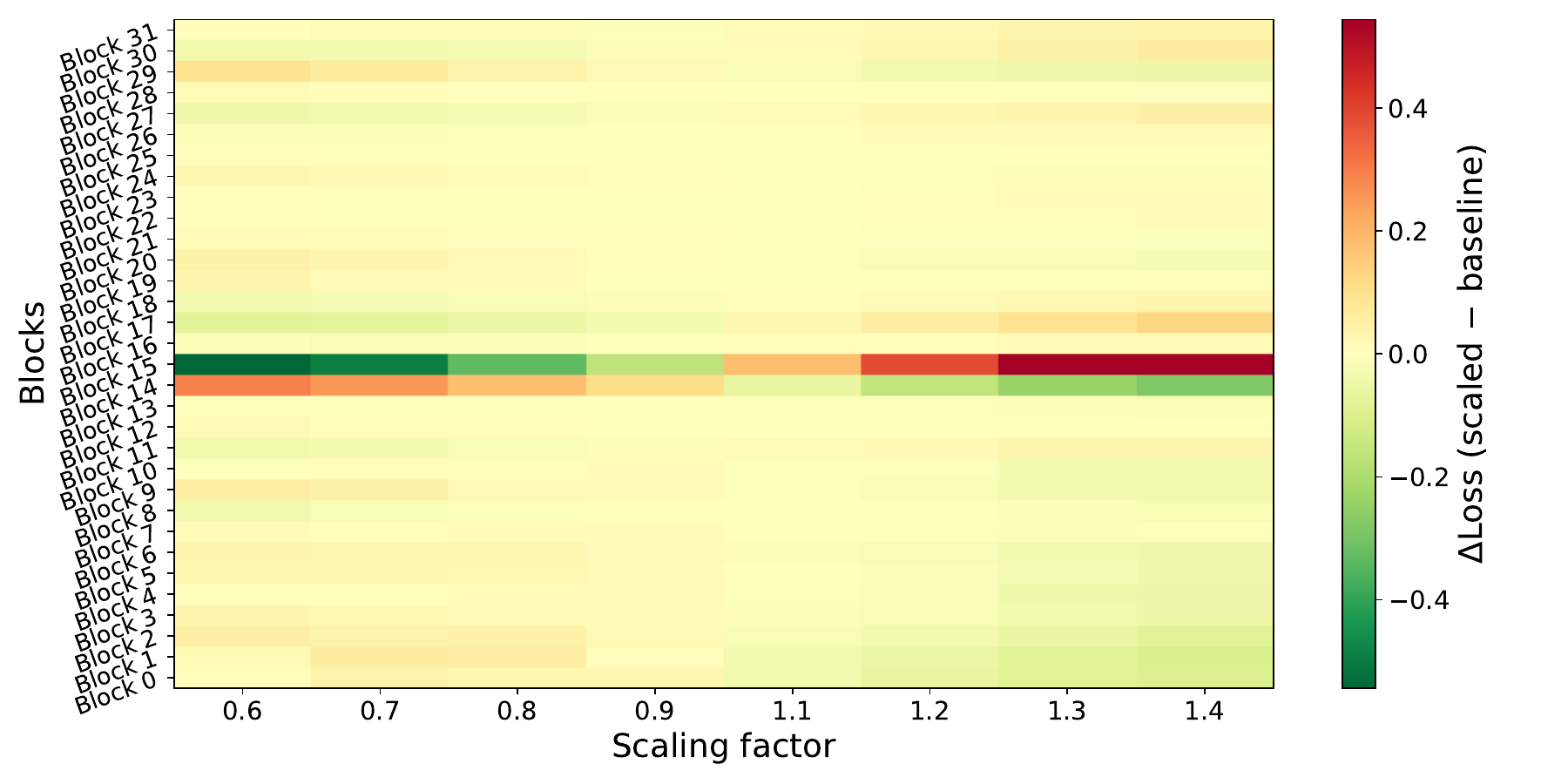}
\caption{}
\end{subfigure}
\hfill
\begin{subfigure}{0.48\linewidth}
\centering
\includegraphics[width=\linewidth]{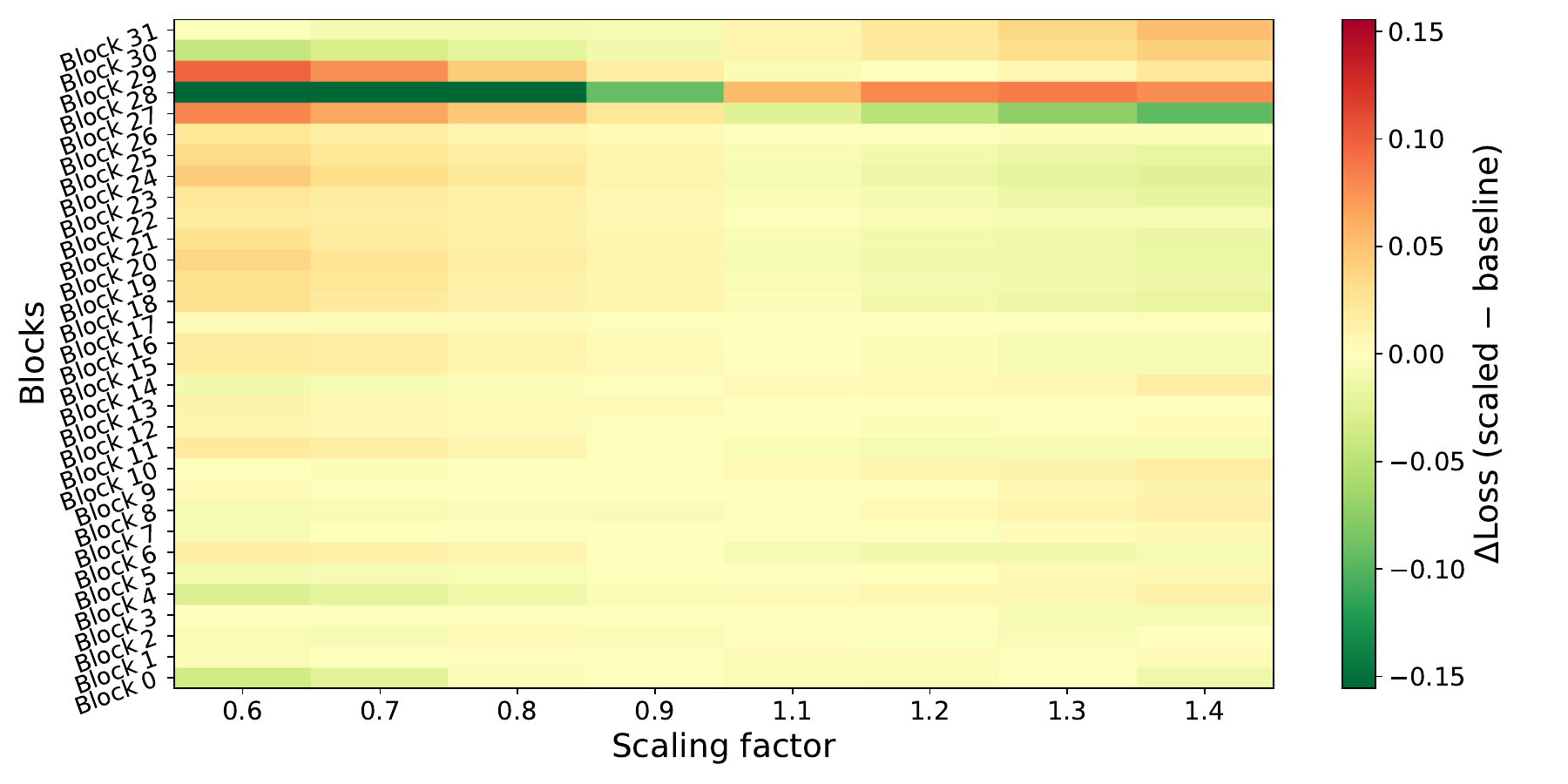}
\caption{}
\end{subfigure}
\caption{Heatmaps of loss change under residual scaling for bit-flips injected into blocks 3, 9, 15, and 28 of LLaMA-3.1 8B. The corrupted block exhibits a distinctive asymmetric loss response, producing the highest Block Sensitivity Score (BSS).}
\label{fig:selfref-8b-bss}
\end{figure*}

\begin{figure*}[h]
    \includegraphics[width=\linewidth]{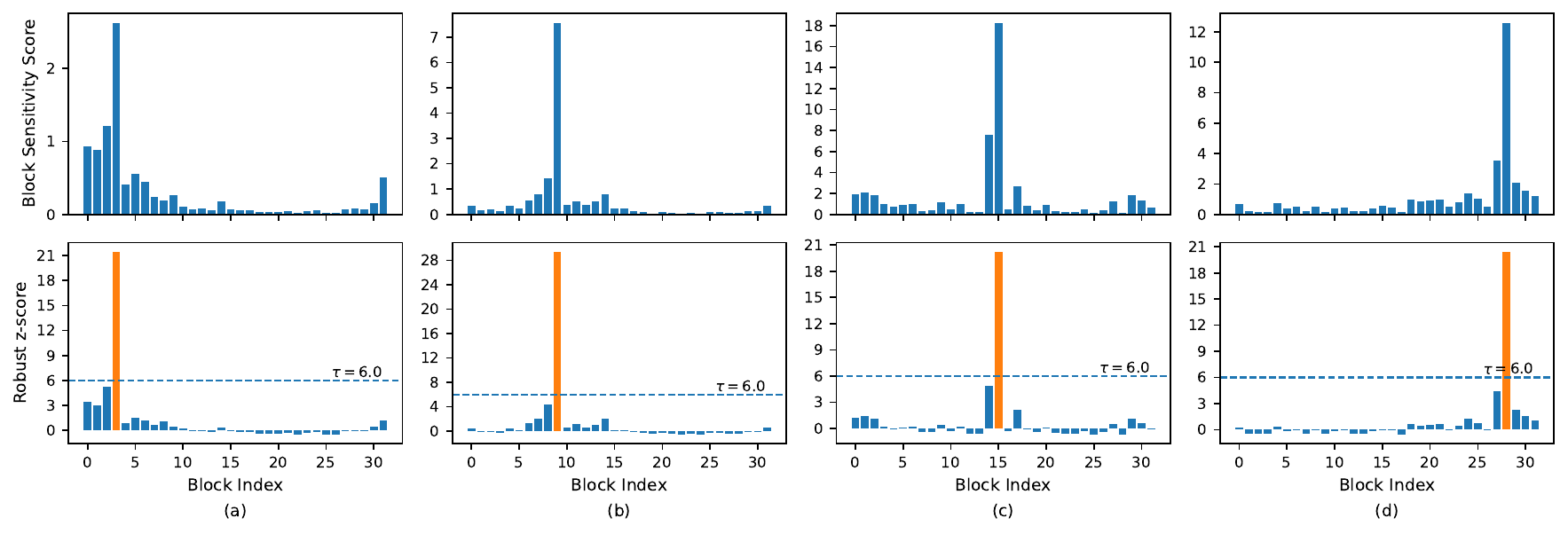}
    \caption{Identification of perturbed blocks in LLaMA~3.1 8B under self-referential configurations, using a threshold based on the BSS and robust z-score computed for each block.}
    \label{fig:results_8B}
\end{figure*}

Fig.~\ref{fig:selfref-8b-bss} shows the corresponding block sensitivity patterns for these injections. The corrupted block consistently exhibits the strongest asymmetric loss response, allowing BitFlipScope to isolate the faulty block. Fig.~\ref{fig:results_8B} summarizes the resulting BSS and robust z-score values across all blocks, confirming that the threshold-based detection reliably identifies the perturbed block.

\end{document}